\documentclass{aa}

\usepackage{graphicx}

\usepackage{txfonts}

\usepackage{hyperref}
\usepackage{amsmath,url}
\usepackage{algorithm}
\usepackage{amsfonts}
\usepackage{mathrsfs}
\usepackage{bm}
\usepackage{rotating}
\usepackage{color}
\usepackage{graphicx}
\usepackage{subfigure}
\usepackage{lineno}
\usepackage[toc,page]{appendix}
\usepackage[english]{babel}

\usepackage{multirow}
\usepackage{graphicx}  
\usepackage{multicol}
\usepackage{array}
\usepackage{tabularx}

\usepackage{longtable}  
\usepackage{graphicx}   
\usepackage{lscape}     
\usepackage{pdflscape}  
\usepackage{tabularx} 
\usepackage{rotating} 
\usepackage{placeins}
\usepackage{tabularx}  
\usepackage{booktabs}  

\usepackage{tabularx}
\usepackage{amsmath} 

\usepackage{threeparttable} 
\usepackage{booktabs}

\usepackage{subcaption}

\usepackage{natbib}
\bibpunct{(}{)}{;}{a}{}{,} 

\begin{document}

   \title{Kinematical and dynamical properties of recently discovered bulge and disc star clusters with WINERED}
   \titlerunning{Short Title}

   \author{Ilaria Petralia\inst{1}
          \and
          Dante Minniti\inst{1,2}
          \and
          Jos\'e G. Fern\'andez-Trincado\inst{3,9}
          \and
          Noriyuki Matsunaga\inst{4,5}
          \and
          Daisuke Taniguchi\inst{5,7}
          \and 
          Sasi Saroon\inst{1}
          \and
          Elisa R. Garro\inst{8}
          \and
          Hiroaki Sameshima\inst{6}
          \and
          Shogo Otsubo\inst{5}
          \and
          Yuki Sarugaku\inst{5}
          \and
          Tomomi Takeuchi\inst{5}
          }

   \institute{Instituto de Astrofísica, Facultad de Ciencias Exactas, Universidad Andrés Bello, Fernández Concha 700, Las Condes,
   Santiago, Chile\\
   \email{ilariapetralia28@gmail.com}
   \and
   Vatican Observatory, Vatican City State, V-00120, Italy
   \and
   Universidad Cat\'olica del Norte, N\'ucleo UCN en Arqueolog\'ia Gal\'actica - Inst. de Astronom\'ia, Av. Angamos 0610, Antofagasta, Chile
   \and
   Department of Astronomy, Graduate School of Science, The University of Tokyo, 7-3-1 Hongo, Bunkyo-ku, 113-0033, Tokyo, Japan
   \and
   Laboratory of Infrared High-resolution spectroscopy (LiH), Koyama Astronomical Observatory, Kyoto Sangyo University, Motoyama, Kamigamo, Kita-ku, 603-8555, Kyoto, Japan
   \and
   Institute of Astronomy, School of Science, The University of Tokyo, 2-21-1 Osawa, Mitaka, Tokyo 181-0015, Japan
   \and
   National Astronomical Observatory of Japan, 2-21-1 Osawa, Mitaka, Tokyo 181-8588, Japan
   \and
   ESO – European Southern Observatory, Alonso de Cordova 3107, Vitacura, Santiago, Chile
   \and
   Universidad Cat\'olica del Norte, Departamento de Ingenier\'ia de Sistemas y Computaci\'on, Av. Angamos 0610, Antofagasta, Chile}

  \abstract
   {Galactic globular clusters are a very important tool in explaining the characteristics of the Milky Way. Therefore it is essential to determine the {kinematical} and dynamical properties of the new star cluster candidates, especially at the low-latitude regions that suffer from heavy extinction and crowding.}
   { In this work, we report the first spectroscopic analysis for seven recently identified star cluster candidates: CWNU 4193, FSR 1700, Garro 02, Patchick 98, FSR 1767, Mercer 08, and BH 140. Our aim is to determine the kinematical properties, such as the {mean cluster radial velocity}, and the dynamical properties, such as the orbital parameters and the global dynamical mass, of these clusters in order to spectroscopically confirm the nature of these
seven stellar systems.}
   {We collected the high-resolution infrared spectra of 33 candidate members of these clusters using the WINERED spectrograph mounted on the Magellan Clay 6.5 m telescope. Using the WINERED spectra, we measured the radial velocity of each individual star to confirm its membership in the clusters. From the confirmed members, we derived the {mean cluster radial velocity} of each cluster. In addition, for these clusters, we computed the {orbital elements} using the \texttt{GravPot16} model and estimated their {global} dynamical masses based on the virial theorem.}
   {As a result, {we confirmed enough member stars (from three to seven stars per cluster) to reliably derive the} {mean cluster radial velocity} and compute the orbital parameters of the clusters CWNU 4193, FSR 1700, Garro 02, and BH 140. For clusters CWNU 4193, FSR 1700, and BH 140, the number of confirmed members also allowed us to estimate their {global} dynamical masses.
   Therefore, we successfully derived key {kinematical and} dynamical properties for four of the most obscured star clusters in the Milky Way.
   }
   {}


   \maketitle

\section{Introduction}\label{intro}

Globular clusters (GCs) are very important tools to study the formation and evolution of their host galaxies, as they are among the oldest objects in the Universe (\citealt{Sagar_1997}). In particular, Galactic GCs are helpful in explaining the characteristics of the Milky Way, { such as its structure and chemical composition since GCs retain the chemical signatures of the early stages of Galactic formation and evolution.}
These objects are also important dynamical probes to investigate dynamic events in the past (\citealt{West_2004}). For these reasons, it is essential to obtain a complete census of these stellar systems.\\
Galactic GCs have been observed and characterised in different environments. However, in our galaxy, there are some regions,
such as the Galactic centre and the Galactic disc, where the observations are limited, due to high extinction effects.\footnote{{The star clusters analysed in this work are located in the regions that were monitored by the VVV survey, namely the Galactic bulge and an adjacent region of the disc. Using the data revealed by the VVV Survey, \citet{2017_Alonso-Garcia} found the mean selective-to-total extinction ratio of $A_{K_{s}}/E(J-K_{s}) = 0.428 \pm 0.005 \pm 0.04$ in the inner Galaxy.}} These environments, in addition to being highly extinct areas, are also very crowded sites.
Nevertheless, over the past few years, infrared surveys such as the Two Micron All-Sky Survey (2MASS,\citealt{2MASS}) and the VISTA Variables in the Via Lactea survey (VVV) and its recent extension the VVVX survey (\citealt{MINNITI_2010}, \citealt{Saito_2024}) have unveiled
new star cluster candidates {($\sim$ 300 GC candidates as reported in \citealt{2024_Garro})} in these regions of our Galaxy. \\
Two of these newly identified stellar systems are the star clusters CWNU 4193 and FSR 1700, both discovered in the Galactic disc.
The star cluster CWNU 4193, located at the J2000 equatorial coordinates $\alpha$ = 08:04:41.7, $\delta$ = -38°:55:16, was discovered by \citet{He_2023}. Recently, utilising the VVVX survey data, \citet{Saroon_2024} studied this cluster photometrically and determined a cluster age of 11 Gyr, a metallicity of [Fe/H] $= -0.85 \pm 0.20 $ dex, a {heliocentric distance} of d = 12.8 $\pm$ 0.5 kpc, and a concentration parameter of C = 1.49. Based on these results, CWNU 4193 emerges as a GC candidate.
On the other hand, the star cluster FSR 1700,
situated at the J2000 equatorial coordinates
$\alpha$ = 15:38:52.5, $\delta$ = -59°:16:03, was discovered by \citet{Froebrich_2007} and classified as a distant reddened cluster by \citet{Buckner_&_Froebrich_2013}. \citet{He_2023} have recently identified this object as a candidate GC, while \citet{Saroon_2024}, using the VVVX survey data, derived a cluster age of 11 Gyr, a metallicity of [Fe/H] = -0.80 $\pm$ 0.20 dex, a {heliocentric distance} of d = 10.3 $\pm$ 0.5 kpc, and concentration parameter of C = 1.33. From these latest parameters, FSR 1700 also emerges as a GC candidate. \\
Two more examples of new candidate star clusters are Garro 02 and FSR 1767, both identified in the Galactic bulge. The Garro 02 cluster, positioned in the Milky Way bulge at the J2000 equatorial coordinates $\alpha$ = 18:05:51.1, $\delta$ = -17°:42:02, was discovered and photometrically studied in \citet{Garro_2022_Garro2}. Using
the VVVX database, this work derived a cluster age of 12 $\pm$ 2 Gyr, a metallicity of [Fe/H] = -1.30 $\pm$ 0.20 dex, a {heliocentric distance} of d = 5.6 $\pm$ 0.8 kpc, and a concentration parameter of C = 1.74, confirming this cluster as a new genuine Galactic GC. A particular case is the FSR 1767 cluster, located at the J2000 equatorial coordinates  $\alpha$ = 17:35:44.8, $\delta$ = -36°:21:42. It has been the subject of several studies in the literature (\citealt{Froebrich_2007}, \citealt{Bonatto_2007}, \citeyear{Bonatto_2009}, \citealt{Buckner_&_Froebrich_2013}, \citealt{Garro_2022_FSR1767}), and its nature has been much debated. \citet{Buckner_&_Froebrich_2013} classified FSR 1767 as an
open cluster, while \citeauthor{Bonatto_2007} (\citeyear{Bonatto_2007}, \citeyear{Bonatto_2009}) and \citet{Garro_2022_FSR1767} classified FSR 1767 as a GC. Each of these studies derived a different value for the {heliocentric distance} and metallicity of this cluster based on the available data and the conclusions drawn. In this work, we adopt the most recent values reported in \citet{Garro_2022_FSR1767}, namely, a
cluster age of 11 $\pm$ 2 Gyr, a {heliocentric distance} of d =  10.6 $\pm$ 0.2 kpc, and a metallicity of [Fe/H] = -0.7 $\pm$ 0.20 dex.\\
Two other star clusters recently identified but still poorly studied are Patchick 98 and Mercer 08. The Patchick 98 cluster is located at the J2000 equatorial coordinates $\alpha$ = 18:18:24 and $\delta$ = -12°:30:06. This star cluster was first identified by amateur astronomer Dana Patchick from Los Angeles, member of the Deep Sky Hunters Collaboration, along with other clusters such as Patchick 122, 125, and 126 that were thoroughly studied by \citeauthor{Garro_Pat122Pat125Pat126} (\citeyear{Garro_Pat122Pat125Pat126}, \citeyear{2023Garro_GaiaIngrins}). The Mercer 08 cluster, situated at the J2000 equatorial coordinates $\alpha$ = 18:28:49 and $\delta$ = -10°:55:55, 
was discovered by \citet{Mercer_2005},
and its nature is still debated.\\
Finally, another recently discovered GC candidate is BH 140. This cluster is situated at the J2000 equatorial coordinates $\alpha$ = 12:53:00.3 and  $\delta$ = -67°:10:28, at a {heliocentric distance} of 4.81 $\pm$ 0.25 kpc. In the past, it was classified as an open cluster (e.g. \citealt{Kharchenko+13}), but in the years that followed, \citet{Cantat-Gaudin+18} provided strong evidence that it is actually a GC. Subsequently, \citet{Soubiran+18} also confirmed that BH 140 is a GC, and it has been included in the catalogues studied by \citet{Vasiliev-Baumgardt+21} and \citet{Belokurov+24}.\\

The present work is structured as follows. In Section \ref{data}, we
present the information about the spectroscopic data of the 33 {member star candidates belonging to the seven clusters}. Section \ref{analysis} reports a spectroscopic analysis with the measurement of the 
radial velocities ($\mathrm{RVs}$).
The results are summarised and discussed in Section \ref{results}. Section \ref{mean_radial_velocity_cluster} shows the calculation of the {mean cluster radial velocity $\left(\overline{\mathrm{RV}}\right)$} of the clusters, while Section \ref{orbit} describes the determination of the {orbital parameters of the clusters.}
In Section \ref{mass_cluster}, we present preliminary mass estimates of the {star clusters}. Finally, our conclusions are drawn in Section \ref{conclusions}.

\section{Spectroscopic data}\label{data}
{In this paper, we present the first spectroscopic analysis to characterise the kinematical and dynamical properties of these seven star cluster candidates. We collected a total of 33 stars: seven stars belonging to CWNU 4193; seven stars belonging to FSR 1700; three stars belonging to Garro 02; four stars belonging to Patchick 98; three stars belonging to FSR 1767; two stars belonging to Mercer 08; and seven stars belonging to BH 140.
The spectra of these stars were acquired (PI: Ilaria Petralia) using the high-resolution near-IR WINERED spectrograph\footnote{\url{http://lihweb.kyoto-su.ac.jp/WINERED/overview_winered.html}} attached to the Magellan Clay 6.5 m telescope at Las Campanas Observatory (LCO) in Chile.}\\
The WINERED spectra covers wavelengths from 0.90 to 1.35 µm ($z'$, Y, and J bands). The spectra were acquired using the 100-µm slit and with a spectral resolution of R = $\lambda/\Delta\lambda$ = 28000 in the WIDE mode (\citealt{Ikeda_2022}).
Information on the date and time at which the WINERED spectra of the 33 stars were acquired is shown in Table {\ref{Table2} (see Appendix \ref{App_table_1}).}
The raw spectral data were processed with the WINERED Automatic Reduction Pipeline (\citealt{Hamano_2024}), resulting in one-dimensional spectra from individual exposures.\\

The 33 stars observed are red giant branch (RGB) stars. The stars have been carefully selected to be likely member stars of the listed clusters{ using a selection based on three criteria: proper motion, astrometry, and magnitude of the stars. As a first step, since these stellar systems are located in high-density regions, a thorough decontamination procedure had to be performed to eliminate stars not belonging to these clusters. A sample of the relatively contaminant-free catalogue of the most probable cluster members has been distilled. It is drawn from the precise astrometry and proper motion from Gaia EDR3 data (\citealt{Gaia_edr3}) and combines VVV and Gaia catalogues.
Using the catalogue of stars selected based on proper motions and astrometry, we further refined the sample by applying an additional selection based on the J-band magnitudes of the stars.
We selected the brightest RGB stars, whose J-band magnitudes are brighter than the red clump of
the clusters in order to facilitate observations.}
Table \ref{Table2} shows the target names, the membership cluster, the coordinates, the J-band magnitudes, {the date and the time of the observations, the heliocentric correction,} the signal-to-noise ratio (S/N) values of the WINERED spectra, the echelle orders used to derive the RV of each star, and the RVs of the 33 observed stars. Note that the S/N values change from the Y-band to the J-band. Indeed, as shown in \citet{Minniti_2024}, where the WINERED spectra of two stars belonging to the VVV CL002 cluster were analysed, for reddened stars, the S/N tends to be significantly low, especially in the Y band.

\section{Measurement of radial velocities}\label{analysis}
\begin{figure}[htbp]

  \centering
  \includegraphics[width=\linewidth]{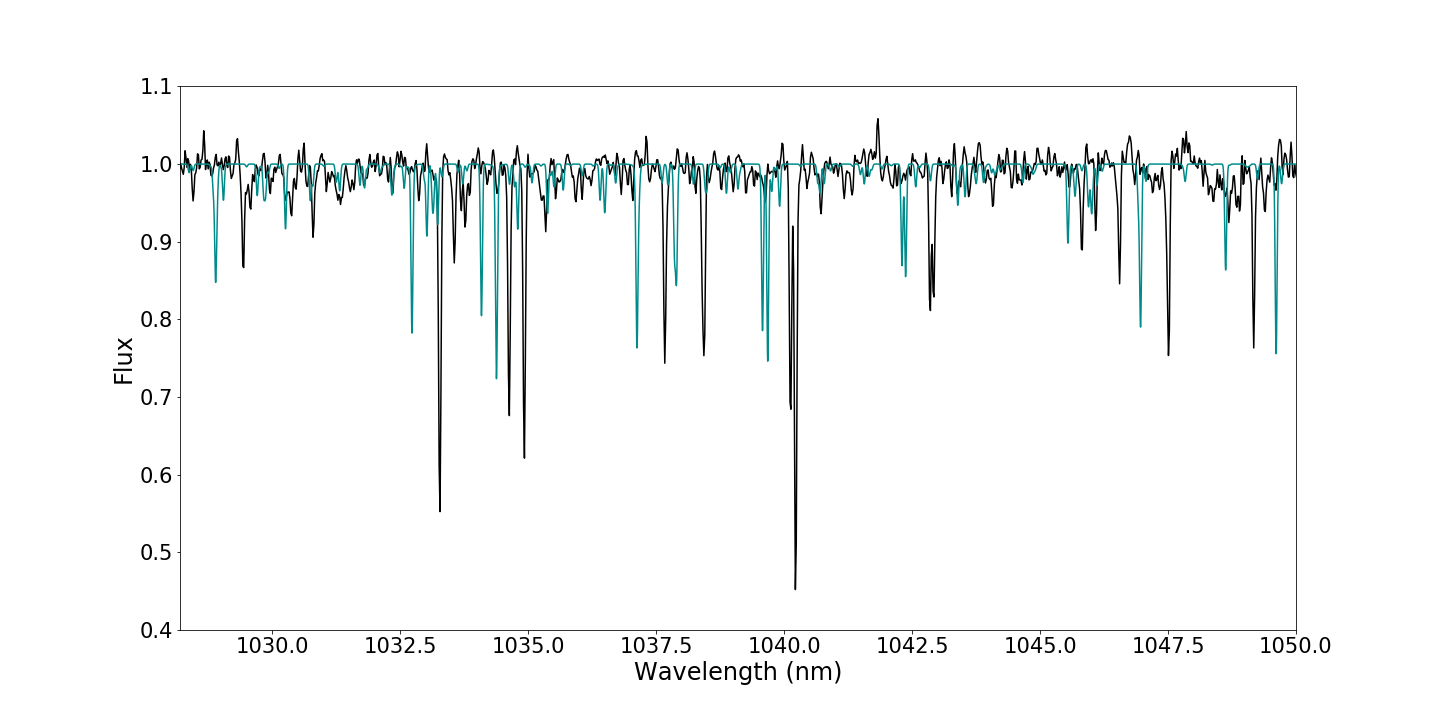}\\

  \caption{WINERED spectra at the 54th echelle order of star 5534905976200827776 (black line). The dark cyan line is the synthetic spectrum, creating with the same atmospheric parameters as Arcturus and adopting for the cross-correlation to derive the {RV.} 
  }
  \label{fig:spectra_antes_despues_HC}  
\end{figure}

The spectroscopic analysis for measuring the RVs 
of the 33 stars belonging to these star clusters candidates was carried out using their spectra in the air wavelength scale, automatically normalised by the pipeline, with each echelle order covering 1.30 times the free spectral range. 
\\
To calculate the RVs of the stars, we adopted the cross-correlation method between the observed spectra and a synthetic spectrum (Figure \ref{fig:spectra_antes_despues_HC}) using the \texttt{iSpec} tool\footnote{\url{https://www.blancocuaresma.com/s/iSpec}}(\citealt{Blanco_Cuaresma_2014}, \citeyear{Blanco_Cuaresma_2019}). We created a synthetic spectrum, adopting the same atmospheric parameters as Arcturus\footnote{The synthetic spectrum has to be similar to the observed spectra, i.e. it has to be created with similar atmospheric parameters of the observed stars. Since the 33 stars are all RGB stars {and their atmospheric parameters are not known}, we decided to create a synthetic spectrum considering the well-known parameters of the RGB star Arcturus. {Indeed, as an RGB star brighter than the red clump, Arcturus has atmospheric parameters similar to those expected for the 33 stars and can thus be adopted as a synthetic spectrum for the cross-correlation.} For the cross-correlation, 
it is possible to use this synthetic spectrum even if the stars observed are more metal rich or more metal poor than Arcturus.} (namely $T_{eff}$= 4286 $\pm$ 35 K, log g = 1.64 $\pm$ 0.06, [M/H]= -0.52 $\pm$ 0.08 dex and macroturbulence = 5 km/s from \citealt{Kondo_2019}), and a spectral resolution of 28,000 (the same resolution of the observed spectra). This was generated using the code SYNTHE (\citealt{Kurucz_1993}), the Kurucz model atmospheres ATLAS9 (\citealt{castelli_2003}), the solar abundances from \citet{Asplund_2005}, and the Vienna Atomic Line Database 3 (VALD3, \citealt{Ryabchikova_2015}). We used this synthetic spectrum for the cross-correlation with each star.\\
An important consideration is that the WINERED spectra of all stars are contaminated by telluric lines. However, the cross-correlation has to be done in spectral regions of the WINERED spectra, where these lines are absent or very weak. As shown in Figure 4 in \citet{Sameshima_2018}, one of the regions without telluric lines ranges from 1.01 to 1.06 µm, i.e. the 55th, 54th, and 53rd echelle order in the WINERED spectra. Thus, for most of the stars, we carried out the cross-correlation in each of these three orders,\footnote{It would have been sufficient to consider only one order to perform the cross-correlation and derive the RV of the star since the offset is the same throughout the spectra. We decided to consider several orders to improve the accuracy of our results.} and then the value of the RV for each star was given by the average of the three values. 
However, for a few stars, we used the 47th and 46th echelle orders (from 1.180 to 1.232 µm, \citealt{Ikeda_2022}) because their entire spectra appear noisy (as shown in Table \ref{Table2}, the noise is much higher than the signal). Thus, in these cases, we chose{d} a region with strong absorption lines, where the probability that the offset is calculated from the star's lines is higher compared to other regions of the spectra, despite the fact that there are telluric lines in this region. Given the difficulty in finding good regions for the calculation of RV for these stars, we derived their final value using only these two orders. Only in two cases (the star 5858116936873419136 and the star 5858116657690170624) were we forced to use a different combination of orders, as one of the orders was not reliable.
The echelle orders used to derive the {RV} of each star are listed in Table \ref{Table2}.\\
Finally, we calculated the heliocentric correction\footnote{The heliocentric correction is an adjustment made to astronomical data to account for the motion of the Earth around the Sun. 
} using the \texttt{rvcorrect} task within IRAF\footnote{IRAF is the Image Reduction and Analysis Facility, a general purpose software system for the reduction and analysis of astronomical data. The software was written by the National Optical Astronomy Observatories (NOAO) in Tucson, Arizona (\url{https://iraf-community.github.io/}).} {considering the information and the coordinates of the stars reported in Table \ref{Table2}.} The heliocentric correction calculated for each star {and the final values of the RV obtained for the 33 stars after the heliocentric correction are shown in Table \ref{Table2}.}\\

The uncertainties of the RVs calculated in the several echelle orders were derived using the \texttt{iSpec} tool, which follows \citet{Zucker_2003}. Considering the propagation of the errors, since the RV value of each star is given by the average of two or three values obtained from the different orders, its uncertainty is given by equation \ref{equazione_errore}:

\begin{equation}
\mathrm{\delta_{\mathrm{RV}} = \mathrm{\frac{1}{N}} \cdot \sqrt{\delta_{\mathrm{RV_1}}^{2}+...+\delta_{\mathrm{RV_N}}^{2}}}
\label{equazione_errore},
\end{equation}

\noindent where N is the number of the echelle orders considered to derive the RV and the addends in the sum in quadrature are the errors of the RVs derived from each individual echelle order. Note that the heliocentric corrections do not have uncertainties (see Table {\ref{Table2}}), and thus the uncertainties of the final values of the RVs of the 33 stars after the heliocentric correction are the same as before the heliocentric correction and are derived from the equation \ref{equazione_errore}.

\section{Results and Discussion}\label{results}
Using the high-resolution infrared WINERED spectrograph, we collected a total of 33 spectra of stars belonging to the target clusters, distributed as follows: seven spectra from CWNU 4193, seven from FSR 1700, three from Garro 02, four from Patchick 98, three from FSR 1767, two from Mercer 08, and seven from BH 140.
From these WINERED spectra, we measured the RVs of the 33 stars, which we subsequently used to derive the {$\overline{\mathrm{RV}}$}
of the clusters, their orbits, and a preliminary estimate of their masses. \\

\subsection{Confirmation of cluster membership}
\begin{figure}
    \centering

    \begin{subfigure}
        \centering
        \includegraphics[width=1\linewidth]{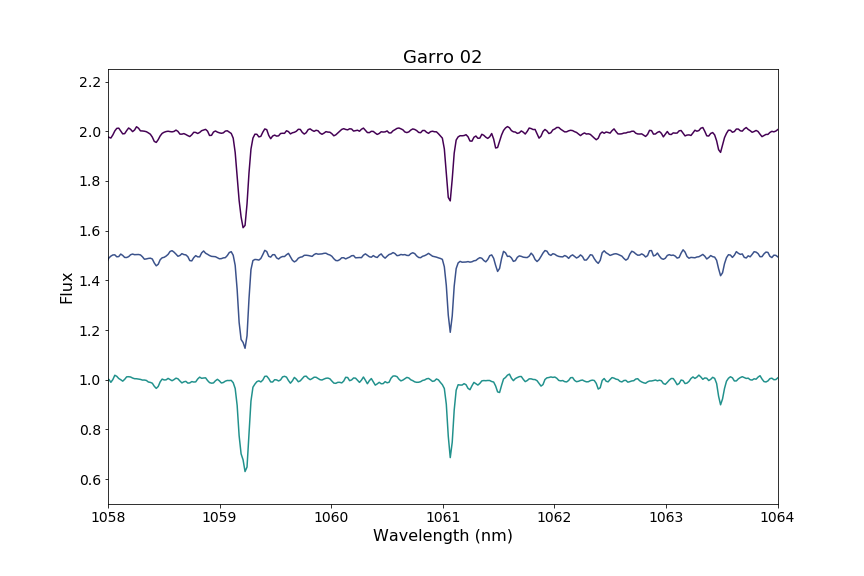}
    \end{subfigure}
    
    \vspace{0.01mm}

    \begin{subfigure}
        \centering
        \includegraphics[width=1\linewidth]{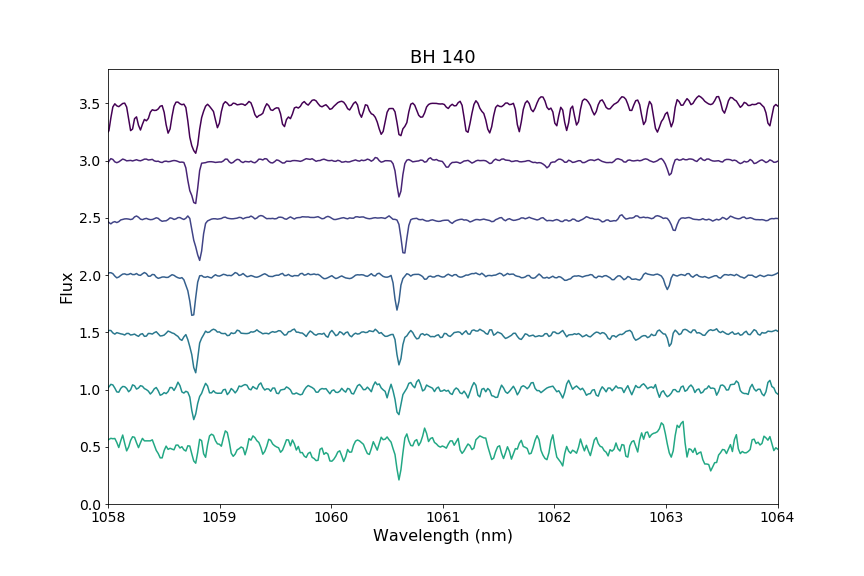}
    \end{subfigure}

    \caption{WINERED spectra of the three member stars of Garro 02 (top) and of the seven member stars of BH 140 (bottom). In the plots, we have shifted the relative flux scale of the stars vertically with additive constants, and we show zoomed spectra between 1058 nm and 1064 nm in the Y-band (53th echelle order). The spectra without corrections are presented, and they show that member stars have lines at the same wavelengths, thus indicating similar {$\mathrm{RVs}$.}
    }
    \label{fig:spettri_Garro02}
\end{figure}

The RVs obtained in this study enabled us to determine whether a star is a member of the reference cluster. To confirm cluster membership, we adopted the 3$\sigma$ rejection criterion. This criterion implies, in fact, that the star has a 99.7\% probability of belonging to the normal velocity distribution of the cluster. Thus, a star is considered a member if its RV lies within 3$\sigma$ of the {$\overline{\mathrm{RV}}$ of the cluster}, namely, 
\begin{equation}
    \mathrm{\left|\mathrm{RV} - \overline{\mathrm{RV}}\right| \leq 3 \cdot \sqrt{\sigma^{2}_{int}+\delta^{2}_{\mathrm{RV}}}},
\end{equation}

\noindent where RV is the radial velocity of the star, $\overline{\mathrm{RV}}$ is the {mean cluster radial velocity}, $\delta_{\mathrm{RV}}$ is the uncertainty of the {radial velocity} of the star, and $\sigma_{int}$ is the intrinsic velocity dispersion 
(the derivation of this parameter is presented in Appendix \ref{App_intrisic_dispersion}).\\
According to this criterion, we found that star 5540909893796469632, a candidate member of cluster CWNU 4193, exhibits a velocity that lies beyond 3$\sigma$ from the cluster mean. Therefore, we excluded this star from our analysis. The same applies to star 5881862471040696320, a candidate member of the cluster FSR 1700. In this case as well, the star was excluded from our analysis. This criterion is useful for excluding candidate stars whose {$\mathrm{RVs}$} 
are significantly different from the {$\overline{\mathrm{RV}}$}
of the reference cluster. However, for stars with {RVs}
close to the cluster’s mean, this criterion can be limiting. In fact, the parent population of GCs typically spans a broader range of {RVs}
than the one considered by this criterion. For this reason, although the 3$\sigma$ rejection criterion indicates that star 5858112023430585216 (candidate members of cluster BH 140) is not a member, we decided not to exclude this star from the analysis and to consider it as member of the cluster since its RV is close to the {$\overline{\mathrm{RV}}$} of the cluster. Moreover, all stars were selected considering their proper motions and spatial positions taken from Gaia DR3 (\citealt{Gaia_DR3}), two additional indicators for establishing cluster membership. {Since this star exhibits proper motion and spatial position consistent with that of its respective parent cluster further supports its classification as member of the cluster.}\\
The other stars belonging to the CWNU 4193 cluster have similar RVs, confirming their membership in this cluster. The same applies to the other stars belonging to the FSR 1700 cluster and the BH 140 cluster. Also, the stars belonging to the Garro 02 cluster show similar RVs, validating their membership in this cluster (see Figure \ref{fig:spettri_Garro02}). Figure \ref{fig:spettri_Garro02} clearly shows that the positions of the spectral lines are aligned across the member stars, indicating that their {RVs}
are comparable or nearly identical. 
Therefore our analysis, which includes the estimation of the {$\overline{\mathrm{RV}}$}, orbit, and mass, will focus on the characterisation of these four clusters.\\
Conversely, for the other three clusters (Patchick 98, FSR 1767, and Mercer 08), it was not possible to determine their {$\overline{\mathrm{RV}}$},
orbit, and mass because the stars identified as cluster members exhibit discrepant velocities, preventing confirmation of their membership. {However, for these three clusters, we performed a qualitative analysis considering the different RV values of their candidate member stars. The results of this analysis are reported in Appendix \ref{speculative_analisi}.} We emphasise that these clusters are located in highly reddened and crowded regions, making it difficult to detect, observe, and confirm their member stars. Therefore, we cannot exclude any star as a member. In order to verify their membership status and to be able to calculate the {$\overline{\mathrm{RV}}$}, orbit, and mass of these clusters, we need additional observations.

\subsection{General considerations and comparison with the literature}

In this section, we briefly present two considerations on the {RV}
values obtained from the WINERED spectra. 
In the case of star 5534906319798217984, due to the S/N being very low, another method that can be adopted for  measuring the RV from the two echelle orders is through equation \ref{formula}:
\begin{equation}
\mathrm{RV} = \frac{\Delta\lambda}{\lambda} \cdot c
    \label{formula},
\end{equation}
where $\Delta\lambda$ is the difference between two centroids (one of the observed spectra and the other of the synthetic spectrum), $\lambda$ is the wavelength of the first centroid, and c is the speed of light ($3\times10^{8}$ m/s). We applied this equation in the range of wavelengths of the 46th echelle order so that we could compare the results obtained from the cross-correlation with those of this method for this order. Since the latter method is based on the difference between two clearly visible centroids, it could be more accurate. However, the result obtained is similar to that obtained by cross-correlation (124.14 km/s using the equation \ref{formula} versus 123.91 km/s from the cross-correlation). Thus, since the result from the cross-correlation is more consistent with the result derived from the 47th echelle order, in our analysis, we adopted the value derived from the cross-correlation.

The other consideration regarding RVs is that it is possible to compare 12 of them with those derived from Gaia DR3 (\citealt{Gaia_DR3}) and to demonstrate the accuracy of our results. Indeed, Gaia's mission has a dedicated spectrometer for calculating RVs. This Gaia instrument determines the RVs using the Doppler shift technique and derives the RVs from spectra around three isolated calcium lines at 849.8, 854.2 and 855.2 nm.\footnote{For more information on the instrument and the techniques used to derive {RVs,}
see \url{https://spaceflight101.com/gaia/gaia-instrument-information/}.} The RV values from the WINERED spectra and those from Gaia's RV spectrometer are shown in Table \ref{Table2}. The RVs derived from WINERED data and the Gaia instrument are generally consistent with each other within the uncertainties. In some cases, { the difference between the WINERED RV values and the Gaia RV values is slightly greater.} 
This is expected, as Gaia is an optical survey, whereas the WINERED spectrograph observes in the infrared. Because of our higher resolution, this comparison also confirms and highlights the good quality of the RVs measured by Gaia, which are important to measure cluster orbits. Nevertheless, considering the values of the uncertainties, the RVs determined from the WINERED spectra are more accurate. \\

\section{{Mean cluster radial velocity of clusters}}\label{mean_radial_velocity_cluster}
We derived the {$\overline{\mathrm{RV}}$}
of the clusters CWNU 4193, FSR 1700, Garro 02, and BH 140 using the confirmed member stars. Our subsequent analysis (described in this section and the following ones) is based on stars confirmed as belonging to their respective clusters, specifically six stars for the CWNU 4193 cluster, five stars for the FSR 1700 cluster, three stars for the Garro 02 cluster, and seven stars for the BH 140.\\
The {$\overline{\mathrm{RV}}$ values}
were derived by taking the average of the RVs of the stars belonging to each cluster, {and the results are reported in Table \ref{table4}}.
However, their uncertainties were derived from the quadrature sum of the statistical error and the error of the standard deviation (see equation \ref{errore_finale}). This is because the few stars collected and confirmed as member stars represent only a sample population. Therefore, to provide a more accurate value of the uncertainty of the {mean cluster radial velocity ($\delta_{\overline{\mathrm{RV}}}$)},
we combined the two errors, which are not completely independent of each other, for each individual cluster.
The statistical error was calculated using the equation \ref{equazione_errore}, but in this case N is the number of stars used to calculate the {$\overline{\mathrm{RV}}$},
while the addends in the sum in quadrature are the errors
of the RVs of each individual member star:

\begin{equation}
\delta_{stat} = \mathrm{\frac{1}{N}} \cdot 
\sqrt{ \mathrm{\sum\limits_{n=1}^N} \delta_{\mathrm{RV_n}}^{2}}
\label{equazione_errore_statistico}.
\end{equation}

\noindent The error of the standard deviation that accounts for the dispersion velocity is given by 

\begin{equation}
    \delta_{SD} = \mathrm{\frac{1}{N}} \cdot \sqrt{\mathrm{\sum\limits_{n=1}^{N}} \left(\mathrm{RV_{n}} - {\overline{\mathrm{RV}}}\right)^2}
\label{errore_sulla_SD},
\end{equation}

\noindent where N is the number of stars used to calculate the {$\overline{\mathrm{RV}}$},
$\mathrm{RV_{n}}$ represents the radial velocities of each member star, and $\overline{\mathrm{RV}}$ is the average value of the {radial velocity} values of the member stars.\\
Thus, the 
{$\delta_{\overline{\mathrm{RV}}}$}
is given by
\begin{equation}
    \delta_{\overline{\mathrm{RV}}} = \sqrt{\delta_{stat}^2 + \delta_{SD}^2}
    \label{errore_finale}.
\end{equation}

\noindent As a result of this analysis {(see Table \ref{table4})}, the {mean cluster radial velocity} of the star cluster CWNU 4193 is $\overline{\mathrm{RV}}$ = {136.61} $\pm$ {0.49} km/s; for the FSR 1700 cluster, it is $\overline{\mathrm{RV}}$ = {5.67 $\pm$ 0.63} km/s; for the Garro 02 cluster, it is $\overline{\mathrm{RV}}$ = {168.79 $\pm$ 0.85} km/s; and for the BH 140 cluster, it is $\overline{\mathrm{RV}}$ = {91.72 $\pm$ 1.92} {{km/s}}.\\
The {$\overline{\mathrm{RV}}$}
derived for clusters CWNU 4193, FSR 1700, and Garro 02 has not been measured previously in the literature. In contrast, the {$\overline{\mathrm{RV}}$}
of BH 140 has been reported in previous studies, allowing for an external comparison with the present work. In \citet{Soubiran+18}, the derived {mean cluster radial velocity} was based on {Gaia DR2 data of} four stars, and they obtained $\overline{\mathrm{RV}}$ = 90.4 $\pm$ 0.9 km/s. This value is consistent with the result obtained in the present analysis. As noted in \citet{Lim2025}, indeed, some discrepancies likely reflect systematic differences between optical and near-IR spectroscopic analyses. Another study that determined {the mean cluster radial velocity}
of this cluster is that of \citet{Vasiliev-Baumgardt+21}, where they found a $\overline{\mathrm{RV}}$ of 90.29 $\pm$ 0.35 km/s. Also in this case, the data used comes from the Gaia survey (in this case, however, from Gaia EDR3), so the slight difference in the results can be justified by the different wavelength regime considered. \\

\renewcommand{\arraystretch}{1.5} 
\begin{table*}[ht]
    \centering
    \caption{Derived {mean cluster radial velocity, intrinsic velocity dispersion, and mass for each cluster.}}
    \resizebox{\linewidth}{!}{%
    \begin{tabular}{ccccccccccc}
    \hline
    Cluster &RA &Dec&d&$\overline{\mu_\alpha\cdot cos\delta}$&$\overline{\mu_{\delta}}$&$N_{*}$& $\overline{\mathrm{RV}}$ & $\sigma_{int}$ & $r_{t}$& Mass\\
        &[hh:mm:ss]&[dd:mm:ss]&[kpc]&[mas/yr]&[mas/yr]&&[km/s]&[km/s]&[pc]&[$\times$10$^{4}$  M$_{\odot}$]\\
\hline
CWNU~4193 &08:04:41.7&$-$38:55:16&12.8 ± 0.5&$-$0.792 ± 0.025&1.628 ± 0.213& 6& {136.61 $\pm$ 0.49} &{0.55 $\pm$ 0.50}&11.4 $\pm$ 0.6 &
{0.08  $\pm$ 0.15} \\
FSR~1700 &15:38:52.5&$-$59:16:03&10.3 ± 0.5&$-$4.850 ± 0.014&$-$4.030 ± 0.013& 5& {5.67 $\pm$ 0.63} & {0.95 $\pm$  0.48}& 26.9$\pm$1.7 & 
{0.56 $\pm$ 0.57}  \\
Garro~02&18:05:51.1&$-$17:42:02&5.6 ± 0.8&$-$6.07 ± 0.62&$-$6.15 ± 0.75&3& {168.79 $\pm$ 0.85} & {0.91 $\pm$  0.78}& 11.6$\pm$6.5 & -\\
BH~140 &12:53:00.3&$-$67:10:28&4.81 ± 0.25&$-$14.848 ± 0.024&1.224 ± 0.024&7&  {91.72 $\pm$ 1.92} & {3.35 $\pm$ 1.06}&31.0$\pm$1.7 & 
{8.07 $\pm$ 5.13}  \\
\hline
\end{tabular}
}
\tablefoot{{The cluster names and their coordinates are listed in columns 1 - 3. The heliocentric distance is shown in the fourth column, while the proper motions are shown in coloumns 5 - 6. The values listed in columns 1–6 are taken from the literature works indicated in Section \ref{intro}. The number of member stars for each cluster in this work are reported in the seventh column. The {$\overline{\mathrm{RV}}$ is}
shown in in the eighth column, while the {intrinsic velocity dispersion} calculated in this work is shown in the ninth column. The literature
tidal radius is listed in the tenth column, converted to parsecs. In particular, we adopted the tidal radius from \citet{Saroon_2024} for CWNU~4193 and FSR~1700 and from \citet{Garro_2022_Garro2} for Garro~02. In the case of BH 140 cluster, we derived its tidal radius from Gaia DR3 data (see Appendix \ref{app:tidal_radius_bh140}). Finally, the last column shows the estimated mass for each cluster.
}}
\label{table4}
\end{table*}

{Kinematical} properties, such as {the $\overline{\mathrm{RV}}$}, 
allowed us to confirm the existence of the stellar clusters, 
thus confirming, or not, the results of the photometric analysis. Indeed, the presence of the candidate stellar cluster can be verified by comparing its {$\overline{\mathrm{RV}}$}
with that of the surrounding field.
For example, for the position of the FSR 1700 cluster, {the mean disc motion\footnote{{By `disc motion', we refer to the large-scale, collective organised kinematics of the stars and gas comprising the Milky Way’s disc. In detail, in this analysis we adopt{ed} the {mean radial velocity of the disc motion} with respect to the Sun.}}} is -40 km/s. This value is completely different from the {$\overline{\mathrm{RV}}$}
of this cluster that we derived. Therefore, since these velocities are different, the existence of this cluster is confirmed.
Moreover, the {$\overline{\mathrm{RV}}$ of each cluster}
can {provide initial insight into whether our star clusters could be GCs. Indeed, in the first instance, it is possible to compare the {$\overline{\mathrm{RV}}$ values}
of our clusters with those of GCs located in the same spatial region. Similar velocities can serve as a preliminary indication supporting their nature as GCs. Secondly, the {$\overline{\mathrm{RV}}$}
can be used to calculate the orbits of the clusters (see Section \ref{orbit}), from which additional information on their characteristics can be obtained that help define their nature more precisely. For example, the} star cluster CWNU 4193 is located {in a region} where the predicted mean disc motion {yields $\sim$ 135 km/s}. This value is similar to the {$\overline{\mathrm{RV}}$}
of {136.61 $\pm$ 0.49 km/s} that we derived for this cluster. {This similarity suggests that the cluster follows the ordered motion of the Galactic disc. Such kinematic behaviour is typically observed in open clusters.
Consequently,} the CWNU 4193 cluster could be an open cluster. 
{In contrast, the results derived from the orbital calculation indicate that it is more likely a thick disc cluster. This suggests that the CWNU 4193 cluster could be a GC since the thick disc is populated mostly by GCs and a few old open clusters. 

When considering only the {$\overline{\mathrm{RV}}$ values}
of the clusters, we find it is not possible to constrain the nature of the new star cluster candidates in this work. However, these kinematical properties can provide initial information that is useful to subsequently accessing the nature of these clusters. Indeed, to} correctly establish the nature of these clusters, further data and an analysis of chemical abundances are required.\\

\section{{Orbital parameters of clusters}}\label{orbit}
We used the state-of-the-art Milky Way model  \texttt{GravPot16}\footnote{\url{https://gravpot.utinam.cnrs.fr}} to predict the orbital path of the clusters CWNU 4193, FSR 1700, Garro 02, and BH 140 in a steady-state gravitational Galactic model that includes a boxy/peanut bar structure \citep{Fernandez-Trincado2019}. We computed the orbits of the four clusters using \texttt{GravPot16} code, which includes the perturbations due to a realistic (as far as possible) rotating boxy/peanut bar, which fits the structural and dynamical parameters of the Galaxy as best we can according to recent knowledge of the Milky Way. \\
For the orbit computations, we adopted the same model configuration, solar position, and velocity vector as described in \citet{Fernandez-Trincado2019}, except for the {bar pattern speed}
($\Omega_{\rm bar}$), for which we employed the recommended value of 41 km s$^{-1}$ kpc$^{-1}$ (\citealt{Sanders_2019}), and by assuming variations of $\pm$10 km s$^{-1}$ kpc$^{-1}$. 

The considered structural parameters of our bar model (e.g. mass and orientation) are within observational estimations, which lie in the range of 1.1 $\times$10$^{10}$ M$_{\odot}$ and the present-day orientation of 20$^{\circ}$ \citep[value adopted from dynamical constraints, as highlighted in fig. 12 of][]{Tang2018} in the non-inertial frame (where the bar is at rest). The bar scale lengths are $x_0=$1.46 kpc, $y_{0}=$ 0.49 kpc, and $z_0=$0.39 kpc, and the middle region ends at the effective semi-major axis of the bar ${\mathrm{R_{c}}} = 3.28$ kpc \citep{Robin2012}. Our bar model locates the corotation radius ({CR}) at 4.2 kpc for $\Omega_{\rm bar} = $ 51 km s$^{-1}$ kpc$^{-1}$, 5.5 kpc for $\Omega_{\rm bar} = $ 41 km s$^{-1}$ kpc$^{-1}$, and 7.4 kpc for $\Omega_{\rm bar} = $ 31 km s$^{-1}$ kpc$^{-1}$.

For guidance, the Galactic convention adopted by this work is as follows: {the} $X-$axis is oriented towards $l=$ 0$^{\circ}$ and $b=$ 0$^{\circ}$, the $Y-$axis is oriented towards $l$ = 90$^{\circ}$ and $b=$0$^{\circ}$, and the disc rotates towards $l=$ 90$^{\circ}$. The velocity is also oriented in these directions. Following this convention, the Sun's orbital velocity vectors are [U$_{\odot}$, V$_{\odot}$, W$_{\odot}$] = [$11.1$, $12.24$, 7.25] km s$^{-1}$ \citep{Ralph2010}. The model was rescaled to the Sun's galactocentric distance, 8.27 kpc \citep{GRAVITY2021}, and 
the circular velocity at the solar position was rescaled to be $\sim$ $229$ km s$^{-1}$ \citep{Eilers2019}.

The most likely orbital parameters and their uncertainties were estimated using a simple Monte Carlo scheme. For each cluster, an ensemble of one million orbits was computed backwards in time for 2 Gyr under variations of the observational parameters while assuming a normal distribution for the uncertainties of the input parameters (e.g. positions, heliocentric distances, 
{radial velocities,}
and proper motions), which were propagated as 1$\sigma$ variations in a Gaussian Monte Carlo resampling. To compute the orbits of CWNU 4193, FSR~1700, Garro 02, and BH 140, we adopted the 
{$\overline{\mathrm{RV}}$} that we measured from our WINERED spectra (see Section \ref{mean_radial_velocity_cluster} and Table {\ref{table4}}). The absolute proper motions $\overline{\mu_{\alpha}\cdot cos \delta}$ and $\overline{\mu_{\delta}}$ were taken from \citet{Saroon_2024} (for the clusters CWNU 4193 and FSR 1700), \citet{Garro_2022_Garro2} (for the Garro 02 cluster), and \citet{Vasiliev-Baumgardt+21} (for the BH 140 cluster), with an assumed uncertainty of 0.5 mas yr$^{-1}$ for the orbit computations. The heliocentric distance 
was adopted from the aforementioned literature works {(See Table \ref{table4})}.

\begin{figure*}[ht]
    \centering
    \includegraphics[trim={0cm 1cm 0 0cm},clip,width=0.75\textwidth]{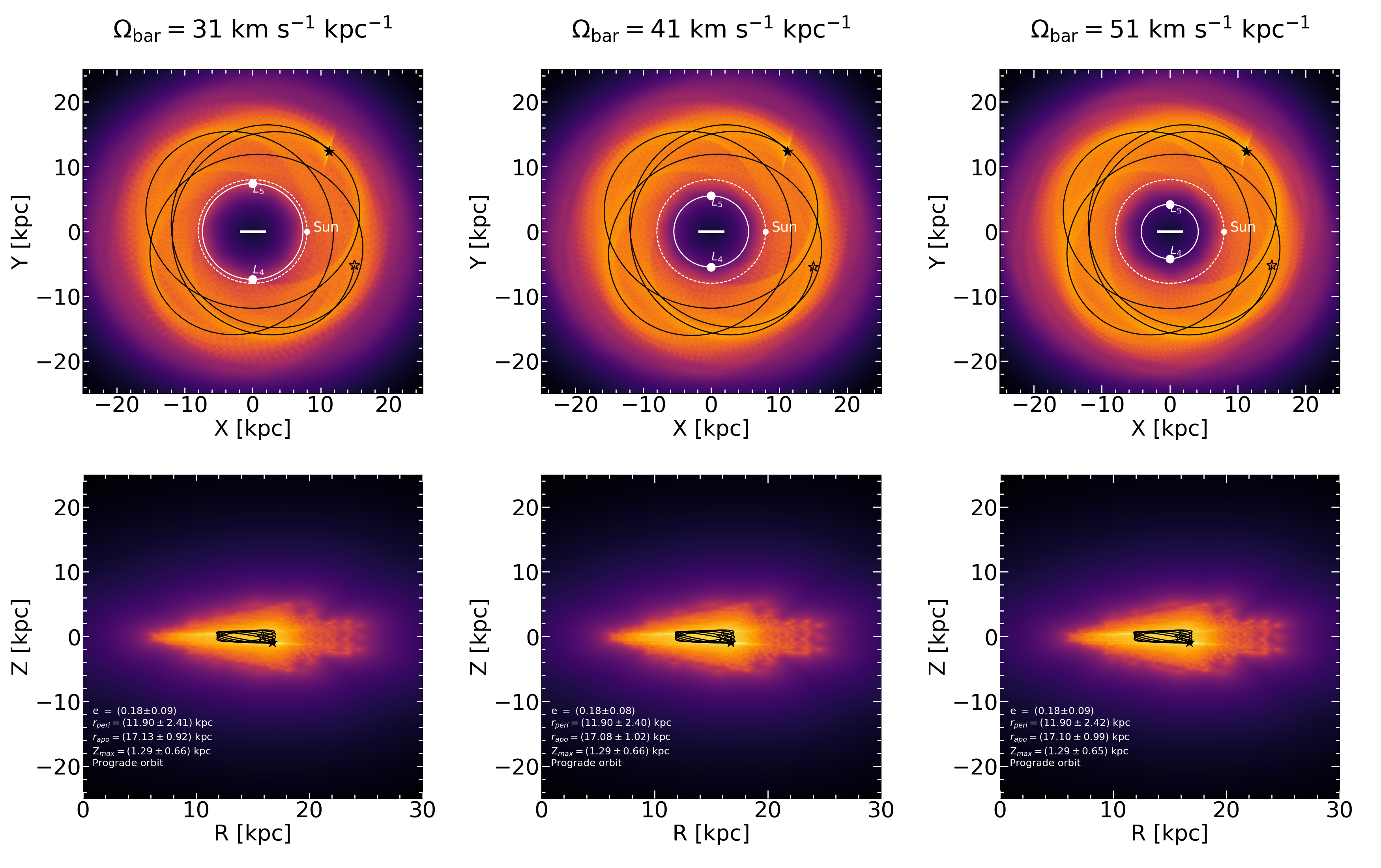}
    \caption{Ensemble of one million orbits in the frame corotating with the bar for the CWNU 4193 cluster, projected on the equatorial ({top}) and meridional ({bottom}) Galactic planes in the non-inertial reference frame with a {$\Omega_{\rm bar}$}
    of 31 ({left}), 41 ({middle}), and 51 ({right}) km s$^{-1}$ kpc$^{-1}$ and time-integrated backwards over 2 Gyr. The yellow and orange colours correspond to more probable regions of the space, which are most frequently crossed by the simulated orbits. The solid inner white circle and the dashed outer circle in the {top} panels show the locations of the {CR} (see text) and the solar orbit, respectively. The white dots mark the positions of the Lagrange points of the Galactic bar, $L_4$ and $L_5$, and the current position of the Sun, respectively. The horizontal solid white line shows the extension of the bar \citep[$\mathrm{R_c} {\sim3.28}$ kpc;][]{Robin2012} in our model. The black-filled and unfilled star symbols indicate the initial and final positions of the cluster in our simulations, respectively. The solid black line shows the orbital path of the CWNU 4193 cluster from the observables without error bars.}
    \label{fig:orbit4193}
\end{figure*}

\begin{figure*}[ht]
    \centering
    \includegraphics[trim={0cm 1cm 0 0cm},clip,width=0.75\textwidth]{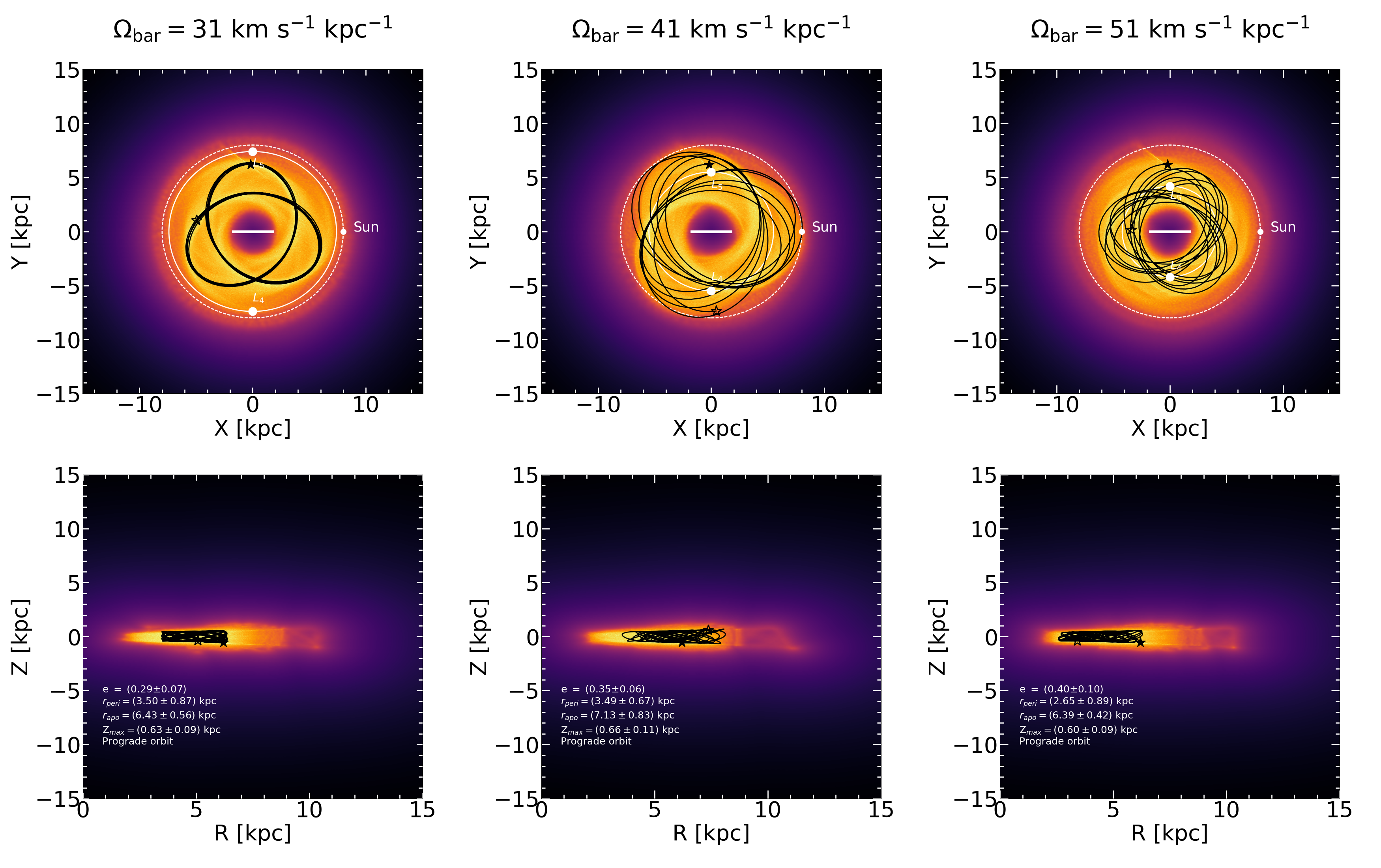}
    \caption{Same as in Figure \ref{fig:orbit4193}, but for the FSR 1700 cluster.}
    \label{fig:orbit1700}
\end{figure*}

\begin{figure*}[ht]
    \centering
    \includegraphics[trim={0cm 1cm 0 0cm},clip,width=0.75\textwidth]{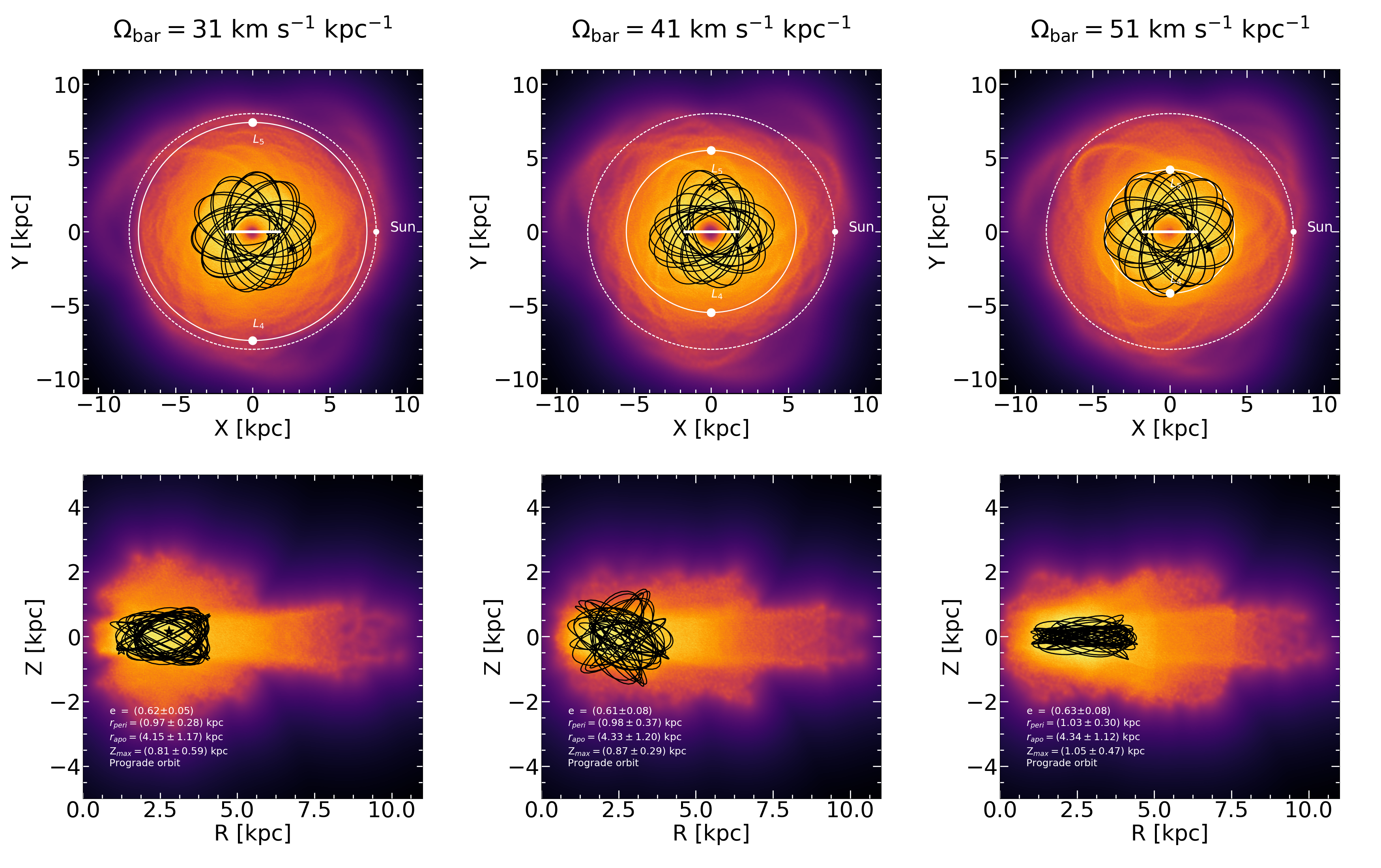}
    \caption{Same as in Figure \ref{fig:orbit4193}, but for the Garro 02 cluster.}
    \label{fig:orbitGarro02}
\end{figure*}

Figures \ref{fig:orbit4193}, \ref{fig:orbit1700}, \ref{fig:orbitGarro02}, and \ref{fig:orbitBH140} show the simulated orbital paths by adopting a simple Monte Carlo approach for the clusters CWNU 4193, FSR 1700, Garro 02, and BH 140, respectively. For each orbit in these figures, we calculate the {perigalactocentric} distance ($r_{peri}$), the {apogalactocentric} distance ($r_{apo}$), the maximum vertical excursion from the Galactic plane {($|Z_{max}|$; hereafter $Z_{\rm max}$)}, and the orbital eccentricity {defined as} $(r_{apo}-r_{peri})/(r_{apo}+r_{peri})$.
The median value of the orbital elements was found for one million realisations, with uncertainty ranges given by the 16$^{\rm th}$ and 84$^{\rm th}$ percentile values, which are listed in the inset of Figures \ref{fig:orbit4193}, \ref{fig:orbit1700}, \ref{fig:orbitGarro02}, and \ref{fig:orbitBH140}. Following the same strategy as described in \citet{Perez-Villegas2018}, we calculated the {z}-component of the angular momentum in the inertial frame
in order to know whether the orbital motion of star clusters has a prograde and/or retrograde sense with respect to the rotation of the bar. Since this quantity is not conserved in a model with bar and/or spiral-arm structures, we were interested only in the sign. Thus the orbital sense for each cluster is also listed in the inset of these figures. The probability densities of the resulting orbits were projected onto the equatorial and meridional galactic planes in the non-inertial reference frame where the bar is at rest. 
In the present integrations, we did not change the other properties of the bar (mass and size). As a caveat, we note that there may be considerable spread in the cluster orbits inferred according to the adopted bar parameters ({$\Omega_{\rm bar}$,}
bar/bulge mass ratio, bar size) for the potentials of different models (\citealt{Smirnov_2024}).
The black line 
{in the Figures \ref{fig:orbit4193}, \ref{fig:orbit1700}, \ref{fig:orbitGarro02}, and \ref{fig:orbitBH140} shows} the orbital path (adopting observables without uncertainties). The yellow colour corresponds to the most probable regions of the space, which are crossed more frequently by the simulated orbits. The orbital elements for the four star clusters are shown in the Table {\ref{tab:Orbital_results}}. In the following subsections we discuss in more detail the individual {dynamical characteristics of the four star clusters}.

{We also note some important limitations of our calculations. We ignored secular changes in the Milky Way potential over time, 
which are expected despite our Galaxy having a quite recent accretion history. We also ignored the fact that all the star clusters were likely affected by dynamical friction, which is expected to bring the clusters closer to the Galactic centre. The bar size was also ignored over different adopted pattern speeds, as this is beyond the scope of this work. Finally, we did not consider perturbations due to spiral arms, as an in-depth analysis is beyond the scope of this paper. We adopted 2 Gyr as the integration time as a conservative limit. For an additional test, we also ran our backwards orbit integration to 1 Gyr instead of 2 Gyr to see how it affects our results. Thus, the overall picture of the orbital properties does not change our main conclusions. On the other hand, the adoption of a longer integration time ($>$ 2 Gyr) will need to account for the above described dynamical effects.
}

\subsection{CWNU~4193}
Figure \ref{fig:orbit4193} shows that CWNU~4193 is confined in a prograde halo orbital configuration, {which lies in a low eccentricity lower than 0.2 with relatively}
low vertical ($Z_{\rm max} \lesssim $ 1.3 kpc) excursions from the Galactic plane, with a {$r_{peri}$}
beyond the solar orbit, at $\sim$ {11.9} kpc, and the largest {$r_{apo}$}
$\lesssim$ {17.13} kpc, which is beyond the boundary of Galactic components. We caught CWNU~4193 near the pericentre of its orbit. {We conclude that this cluster assembly has an orbit more consistent with the thick disc.}

\subsection{FSR~1700}

Figure \ref{fig:orbit1700} shows that FSR~1700 is confined in a prograde inner halo-like orbital configuration, {which lies in an in-plane orbit with low eccentricity $\lesssim$ 0.4 and}
low vertical ($Z_{\rm max} \lesssim $ 0.7 kpc) excursions from the Galactic plane, {$r_{apo}$}
of $\lesssim$ {7.13} kpc, and {$r_{peri}$}
between {2.65 and 3.50} kpc, placing it outside the bulge-bar boundary. Depending on the {$\Omega_{\rm bar}$,}
FSR~1700 has energies to cross the {CR} 
multiple times, but it is confined within the solar orbit. We find that FSR~1700 is not a cluster that lives in the inner bulge region. {Instead, it is more likely a thick disc cluster.}

\subsection{Garro~02}

Figure \ref{fig:orbitGarro02} shows that Garro~02 is clearly confined in a bulge-like orbital configuration that lies in an in-plane orbit with high eccentricity, $\gtrsim$ 0.6, and low vertical ($Z_{\rm max} \lesssim $ {1.05} kpc) excursions from the Galactic plane and {$r_{apo}$}
below {4.35} kpc, which is just below the {CR},
and the {$r_{peri}$}
is $\sim$ 1 kpc. This cluster is both inside and outside of the bar in the Galactic plane,
but it does not have a bar-shape orbit, which means that this cluster is not trapped by the bar, and it is confined to a prograde orbit. The effect produced by different {$\Omega_{\rm bar}$}
on this cluster is almost negligible. We find that Garro~02 is a cluster that lives in the inner bulge region.

\subsection{BH~140}

\begin{figure*}[ht]
    \centering
    \includegraphics[trim={0cm 1cm 0 0cm},clip,width=0.75\textwidth]{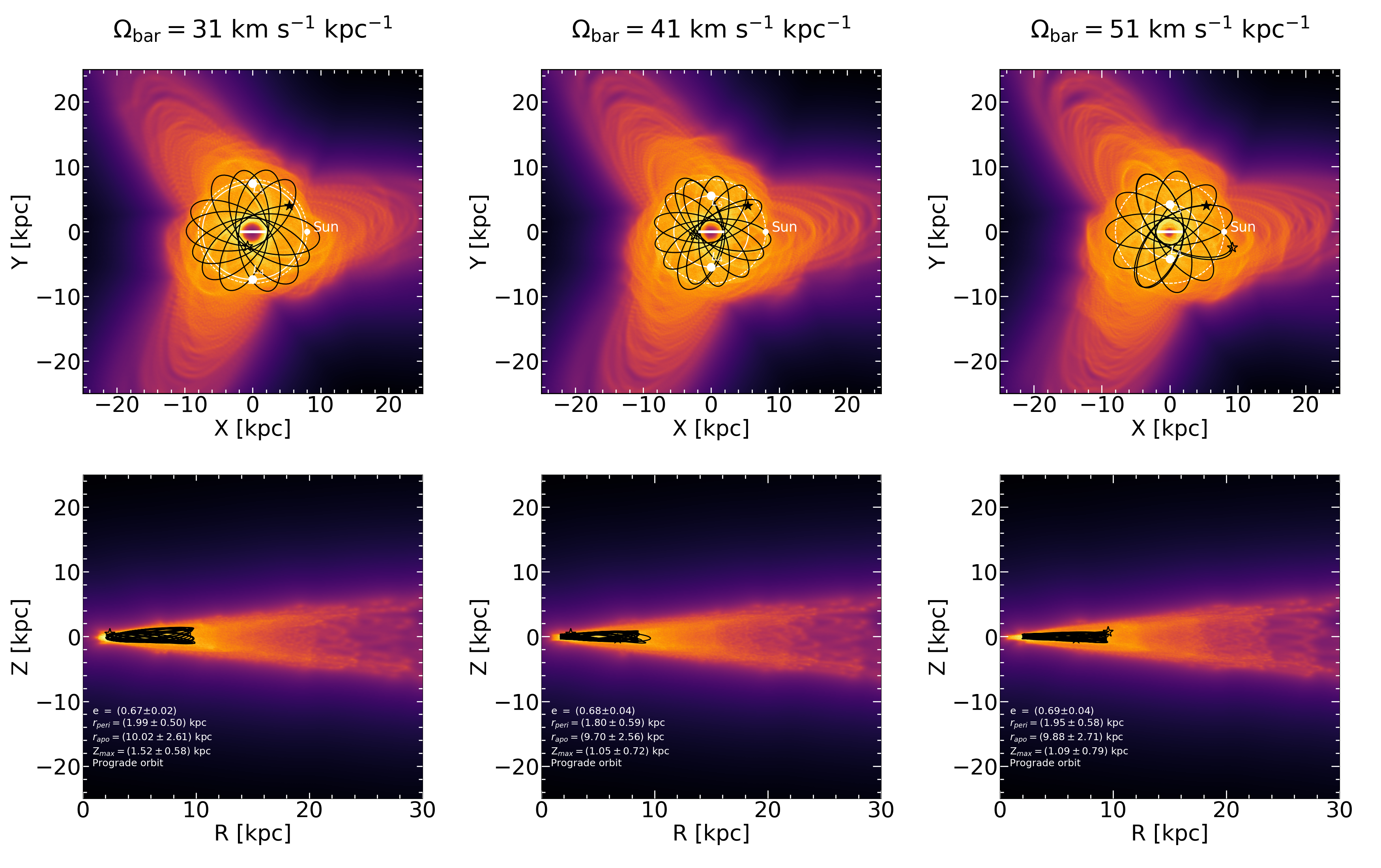}
    \caption{Same as in Figure \ref{fig:orbit4193}, but for the BH 140 cluster.}
    \label{fig:orbitBH140}
\end{figure*}

Figure \ref{fig:orbitBH140} shows that BH~140 is confined in a prograde halo orbital configuration that lies in a high eccentricity, {$\gtrsim$} 0.67, with relatively low vertical ($Z_{\rm max} \lesssim $ 1.52 kpc) excursions from the Galactic plane and with a {$r_{peri}$}
with incursions within the Galactic bulge at $\lesssim$ 1.99 kpc and {$r_{apo}$}
beyond the solar position, $\lesssim$ 10 kpc. 
{We caught BH 140 near the 
$r_{apo}$ of its orbit.}
We conclude that BH~140 appears to be a halo intruder from other stellar components that are currently crossing the central parts of the Galaxy.

\section{{Mass estimates of clusters}}\label{mass_cluster}
Finally, we present the total mass estimate of the clusters using the virial theorem. For a gravitationally bound system in dynamical equilibrium, such as star clusters, the virial theorem relates the average total kinetic energy $\left(\overline{T}\right)$ and the average total gravitational potential energy $\left(\overline{U}\right)$ through the equation
\begin{equation}
    2\cdot \overline{T} + \overline{U} = 0,
\end{equation}

\noindent where
\begin{equation}
    \overline{T} = \frac{1}{2}\cdot \mathrm{M}\cdot \sigma^2_{int} \quad \text{and} \quad
    \overline{U} = -\frac{\mathrm{G}\cdot \mathrm{M}^2}{r_{t}}.
\end{equation}

\noindent Here, M is the {global} dynamical mass, $\sigma_{int}$ is the {intrinsic velocity dispersion}, G is the gravitational constant (6.67 $\times 10^{-11}$ m$^{3}/$kg s$^{2}$), and $r_{t}$ is the tidal radius.
By substituting the equations of the kinetic and the potential energy and solving the formula of the virial theorem to derive the mass, we obtained the following equation:

\begin{equation}
    \mathrm{M} = \frac{ r_{t} \cdot \sigma^{2}_{int} }{\mathrm{G}}.
\end{equation}

\noindent And specifically, when taking into account the units of measurement, we obtained
\begin{equation}
    \mathrm{M} = 2.32 \times 10^{2} \cdot  \left(\frac{r_{t}}{\mathrm{pc}}\right) \cdot \left(\frac{\sigma_{int}}{\mathrm{km/s}}\right)^{2} \quad [\mathrm{M}_\odot].
\end{equation}

\noindent The derivation of these equations is based on the assumptions of an isothermal and isotropic system. The uncertainty on the mass was calculated using error propagation, according to the following formula:

\begin{equation}
    \delta_\mathrm{M} = \mathrm{M} × \sqrt{\left(\frac{\delta_{r_{t}}}{r_t}\right)^2+\left(2\cdot\frac{\delta_{\sigma_{int}}}{\sigma_{int}}\right)^2}.
\end{equation}
The values of the {global} dynamical mass of the clusters are reported in Table \ref{table4}.

{An important consideration in this analysis is the assumption that the star clusters are virialised. As described in the previous sections, all clusters analysed in this work are located in regions characterised by a complex and dynamic gravitational environment. Consequently, these systems could be perturbed by the Galactic tidal field, thus potentially limiting the applicability of the virial theorem for estimating their masses. 
However, by using the individual {RVs}
of the member stars and the tidal radius, this method provides an approximate estimate of the 
global mass of each cluster.
Accordingly,
the masses derived in this analysis should be regarded as preliminary estimates. Moreover, in order} to achieve more accurate values, it is essential to increase the number of member stars analysed for each cluster. Nevertheless, these estimates provide a first indication of the cluster masses.
In the case of the Garro 02 cluster, it is not possible to estimate the mass due to the limited number of available member stars. Indeed, in this work, we performed the mass estimates exclusively for clusters with at least five confirmed member stars.\\
\cite{Vasiliev-Baumgardt+21} provides an estimated mass of $6.1 \times 10^4  \mathrm{M}_\odot$ for BH 140. 
{This value is consistent in order of magnitude with} our estimate, differing by only {32}$\%$. Such a small {relative} difference suggests a good level of agreement between the two determinations despite potential variations arising from different methodologies and datasets. This concordance reinforces the reliability of the mass estimates obtained for this cluster.\\
Typically, open clusters have masses ranging from $10^{2}$ to $10^{4}$ $ \mathrm{M}_\odot$, while GCs have masses between $10^{4}$ and $10^{6}$ $ \mathrm{M}_\odot$.
The mass results obtained for all clusters confirm the findings reported in Section \ref{mean_radial_velocity_cluster}. Indeed, the clusters have masses that lie at the boundary between the typical mass ranges of open clusters and GCs. Therefore, these stellar systems could either be massive open clusters or low-mass highly evolved and dissolved GCs.
Therefore, as in the case of {the $\overline{\mathrm{RV}}$ values},
the masses of these clusters alone do not allow us to define a clear criterion to determine their nature. A chemical analysis and the addition of further data are necessary.

\section{Conclusions}\label{conclusions}
The main goal of the present paper was to derive the {kinematical and dynamical} properties of the seven newly found Milky Way star clusters CWNU 4193, FSR 1700, Garro 02, Patchick 98, FSR 1767, Mercer 08, and BH 140. For this work, we collected a total of 33 WINERED spectra of star candidate members of these clusters, from which we determined the individual RVs of these stars (see Table \ref{Table2}). Our results show that highly precise {RVs}
can be derived from WINERED spectra. We also found good agreement for the RVs obtained from WINERED spectra and Gaia DR3 data for 12 individual stars, even if the Gaia RV errors are larger than the WINERED ones, as expected.\\
These remarkable {RVs}
enabled us to confirm the membership of the stars in the clusters CWNU 4193, FSR 1700, Garro 02, and BH 140 and to derive the {$\overline{\mathrm{RV}}$ for each cluster},
%
{which we then used to
reconstruct their orbits using the \texttt{GravPot16} model. In addition, for the clusters CWNU 4193, FSR 1700, and BH 140, the precise RVs of their member stars also allowed us to
provide the first estimates of their global dynamical masses using the virial theorem.} Nevertheless, considering both the {$\overline{\mathrm{RV}}$ 
and the global} dynamical mass of these clusters, it remains unclear whether they are GCs or open clusters.\\
Conversely, for the clusters Patchick 98, FSR 1767, and Mercer 08, the discrepant individual velocities prevented us from confirming membership and  deriving the {$\overline{\mathrm{RV}}$ values},
orbits, and masses of the clusters.
Therefore, considering the results presented in this work, we successfully derived key {kinematical and} dynamical properties for four of the most obscured star clusters in the Milky Way. However, to conclusively define the nature of all seven star cluster candidates, further observations and an analysis of chemical abundances are required.

\begin{acknowledgements}\\

{I.P. acknowledges support from ANID BECAS/DOCTORADO NACIONAL 21230761.}\\
We gratefully acknowledge the use of data from the ESO Public Survey program IDs 179.B-2002 and 198.B-2004 taken with the VISTA
telescope and data products from the Cambridge Astronomical Survey Unit. \\
This paper is supported by JSPS Bilateral Program Number JPJSBP120239909
and based on the WINERED data gathered with the 6.5 meter Magellan Telescope located
at Las Campanas Observatory, Chile. We are thankful to the staff of Las Campanas Observatory for their support during
the WINERED’s installation and observations. We also acknowledge the Chilean National Time Allocation Committee (CNTAC) for allocated us observing time with the WINERED spectrograph/Magellan Telescope at LCO, Chile, under the programme n° CN2025A-58.\\
D.M. gratefully acknowledges support from the Center for Astrophysics and Associated Technologies CATA by the ANID BASAL projects ACE210002 and FB210003, by Fondecyt Project No. 1220724.\\
{J.G.F-T gratefully acknowledges the grants support provided by ANID Fondecyt Iniciaci\'on No. 11220340, ANID Fondecyt Postdoc No. 3230001, from the Joint Committee ESO-Government of Chile under the agreement 2021 ORP 023/2021 and 2023 ORP 062/2023. 
}

\end{acknowledgements}

\bibliographystyle{aa} 
\bibliography{bibliography} 

@ARTICLE{Saroon_2024,
       author = {{Saroon}, S. and {Dias}, B. and {Minniti}, D. and {Parisi}, M.~C. and {G{\'o}mez}, M. and {Alonso-Garc{\'\i}a}, J.},
        title = "{Three new Galactic globular cluster candidates: FSR1700, Teutsch67, and CWNU4193}",
      journal = {\aap},
         year = 2024,
        month = sep,
       volume = {689},
          eid = {A115},
        pages = {A115},
          doi = {10.1051/0004-6361/202450019},
archivePrefix = {arXiv},
       eprint = {2406.09216},
 primaryClass = {astro-ph.GA},
       adsurl = {https://ui.adsabs.harvard.edu/abs/2024A&A...689A.115S}
}

@article{Ikeda_2022,
doi = {10.1088/1538-3873/ac1c5f},
url = {https://dx.doi.org/10.1088/1538-3873/ac1c5f},
year = {2022},
month = {jan},
publisher = {The Astronomical Society of the Pacific},
volume = {134},
number = {1031},
pages = {015004},
author = {Yuji Ikeda and Sohei Kondo and Shogo Otsubo and Satoshi Hamano and Chikako Yasui and Noriyuki Matsunaga and Hiroaki Sameshima and Tomohiro Yoshikawa and Kei Fukue and Kenshi Nakanishi and Takafumi Kawanishi and Ayaka Watase and Tetsuya Nakaoka and Akira Arai and Masaomi Kinoshita and Ayaka Kitano and Kazuki Nakamura and Akira Asano and Keiichi Takenaka and Taichi Murai and Hideyo Kawakita and Atsushi Minami and Natsuko Izumi and Ryo Yamamoto and Misaki Mizumoto and Daisuke Taniguchi and Takuji Tsujimoto},
title = {Highly Sensitive, Non-cryogenic NIR High-resolution Spectrograph, WINERED},
journal = {\pasp}
}

@ARTICLE{Hamano_2024,
       author = {{Hamano}, Satoshi and {Ikeda}, Yuji and {Otsubo}, Shogo and {Katoh}, Haruki and {Fukue}, Kei and {Matsunaga}, Noriyuki and {Taniguchi}, Daisuke and {Kawakita}, Hideyo and {Takenaka}, Keiichi and {Kondo}, Sohei and {Sameshima}, Hiroaki},
        title = "{WARP: The Data Reduction Pipeline for the WINERED Spectrograph}",
      journal = {\pasp},
         year = 2024,
        month = jan,
       volume = {136},
       number = {1},
          eid = {014504},
        pages = {014504},
          doi = {10.1088/1538-3873/ad1b38},
archivePrefix = {arXiv},
       eprint = {2401.04876},
 primaryClass = {astro-ph.IM},
       adsurl = {https://ui.adsabs.harvard.edu/abs/2024PASP..136a4504H}
}

@ARTICLE{Gaia_edr3,
       author = {{Gaia Collaboration} and {Brown}, A.~G.~A. and {Vallenari}, A. and {Prusti}, T. and {de Bruijne}, J.~H.~J. and {Babusiaux}, C. and {Biermann}, M. and {Creevey}, O.~L. and {Evans}, D.~W. and {Eyer}, L. and {Hutton}, A. and {Jansen}, F. and {Jordi}, C. and {Klioner}, S.~A. and {Lammers}, U. and {Lindegren}, L. and {Luri}, X. and {Mignard}, F. and {Panem}, C. and {Pourbaix}, D. and {Randich}, S. and {Sartoretti}, P. and {Soubiran}, C. and {Walton}, N.~A. and {Arenou}, F. and {Bailer-Jones}, C.~A.~L. and {Bastian}, U. and {Cropper}, M. and {Drimmel}, R. and {Katz}, D. and {Lattanzi}, M.~G. and {van Leeuwen}, F. and {Bakker}, J. and {Cacciari}, C. and {Casta{\~n}eda}, J. and {De Angeli}, F. and {Ducourant}, C. and {Fabricius}, C. and {Fouesneau}, M. and {Fr{\'e}mat}, Y. and {Guerra}, R. and {Guerrier}, A. and {Guiraud}, J. and {Jean-Antoine Piccolo}, A. and {Masana}, E. and {Messineo}, R. and {Mowlavi}, N. and {Nicolas}, C. and {Nienartowicz}, K. and {Pailler}, F. and {Panuzzo}, P. and {Riclet}, F. and {Roux}, W. and {Seabroke}, G.~M. and {Sordo}, R. and {Tanga}, P. and {Th{\'e}venin}, F. and {Gracia-Abril}, G. and {Portell}, J. and {Teyssier}, D. and {Altmann}, M. and {Andrae}, R. and {Bellas-Velidis}, I. and {Benson}, K. and {Berthier}, J. and {Blomme}, R. and {Brugaletta}, E. and {Burgess}, P.~W. and {Busso}, G. and {Carry}, B. and {Cellino}, A. and {Cheek}, N. and {Clementini}, G. and {Damerdji}, Y. and {Davidson}, M. and {Delchambre}, L. and {Dell'Oro}, A. and {Fern{\'a}ndez-Hern{\'a}ndez}, J. and {Galluccio}, L. and {Garc{\'\i}a-Lario}, P. and {Garcia-Reinaldos}, M. and {Gonz{\'a}lez-N{\'u}{\~n}ez}, J. and {Gosset}, E. and {Haigron}, R. and {Halbwachs}, J. -L. and {Hambly}, N.~C. and {Harrison}, D.~L. and {Hatzidimitriou}, D. and {Heiter}, U. and {Hern{\'a}ndez}, J. and {Hestroffer}, D. and {Hodgkin}, S.~T. and {Holl}, B. and {Jan{\ss}en}, K. and {Jevardat de Fombelle}, G. and {Jordan}, S. and {Krone-Martins}, A. and {Lanzafame}, A.~C. and {L{\"o}ffler}, W. and {Lorca}, A. and {Manteiga}, M. and {Marchal}, O. and {Marrese}, P.~M. and {Moitinho}, A. and {Mora}, A. and {Muinonen}, K. and {Osborne}, P. and {Pancino}, E. and {Pauwels}, T. and {Petit}, J. -M. and {Recio-Blanco}, A. and {Richards}, P.~J. and {Riello}, M. and {Rimoldini}, L. and {Robin}, A.~C. and {Roegiers}, T. and {Rybizki}, J. and {Sarro}, L.~M. and {Siopis}, C. and {Smith}, M. and {Sozzetti}, A. and {Ulla}, A. and {Utrilla}, E. and {van Leeuwen}, M. and {van Reeven}, W. and {Abbas}, U. and {Abreu Aramburu}, A. and {Accart}, S. and {Aerts}, C. and {Aguado}, J.~J. and {Ajaj}, M. and {Altavilla}, G. and {{\'A}lvarez}, M.~A. and {{\'A}lvarez Cid-Fuentes}, J. and {Alves}, J. and {Anderson}, R.~I. and {Anglada Varela}, E. and {Antoja}, T. and {Audard}, M. and {Baines}, D. and {Baker}, S.~G. and {Balaguer-N{\'u}{\~n}ez}, L. and {Balbinot}, E. and {Balog}, Z. and {Barache}, C. and {Barbato}, D. and {Barros}, M. and {Barstow}, M.~A. and {Bartolom{\'e}}, S. and {Bassilana}, J. -L. and {Bauchet}, N. and {Baudesson-Stella}, A. and {Becciani}, U. and {Bellazzini}, M. and {Bernet}, M. and {Bertone}, S. and {Bianchi}, L. and {Blanco-Cuaresma}, S. and {Boch}, T. and {Bombrun}, A. and {Bossini}, D. and {Bouquillon}, S. and {Bragaglia}, A. and {Bramante}, L. and {Breedt}, E. and {Bressan}, A. and {Brouillet}, N. and {Bucciarelli}, B. and {Burlacu}, A. and {Busonero}, D. and {Butkevich}, A.~G. and {Buzzi}, R. and {Caffau}, E. and {Cancelliere}, R. and {C{\'a}novas}, H. and {Cantat-Gaudin}, T. and {Carballo}, R. and {Carlucci}, T. and {Carnerero}, M.~I. and {Carrasco}, J.~M. and {Casamiquela}, L. and {Castellani}, M. and {Castro-Ginard}, A. and {Castro Sampol}, P. and {Chaoul}, L. and {Charlot}, P. and {Chemin}, L. and {Chiavassa}, A. and {Cioni}, M. -R.~L. and {Comoretto}, G. and {Cooper}, W.~J. and {Cornez}, T. and {Cowell}, S. and {Crifo}, F. and {Crosta}, M. and {Crowley}, C. and {Dafonte}, C. and {Dapergolas}, A. and {David}, M. and {David}, P. and {de Laverny}, P. and {De Luise}, F. and {De March}, R. and {De Ridder}, J. and {de Souza}, R. and {de Teodoro}, P. and {de Torres}, A. and {del Peloso}, E.~F. and {del Pozo}, E. and {Delbo}, M. and {Delgado}, A. and {Delgado}, H.~E. and {Delisle}, J. -B. and {Di Matteo}, P. and {Diakite}, S. and {Diener}, C. and {Distefano}, E. and {Dolding}, C. and {Eappachen}, D. and {Edvardsson}, B. and {Enke}, H. and {Esquej}, P. and {Fabre}, C. and {Fabrizio}, M. and {Faigler}, S. and {Fedorets}, G. and {Fernique}, P. and {Fienga}, A. and {Figueras}, F. and {Fouron}, C. and {Fragkoudi}, F. and {Fraile}, E. and {Franke}, F. and {Gai}, M. and {Garabato}, D. and {Garcia-Gutierrez}, A. and {Garc{\'\i}a-Torres}, M. and {Garofalo}, A. and {Gavras}, P. and {Gerlach}, E. and {Geyer}, R. and {Giacobbe}, P. and {Gilmore}, G. and {Girona}, S. and {Giuffrida}, G. and {Gomel}, R. and {Gomez}, A. and {Gonzalez-Santamaria}, I. and {Gonz{\'a}lez-Vidal}, J.~J. and {Granvik}, M. and {Guti{\'e}rrez-S{\'a}nchez}, R. and {Guy}, L.~P. and {Hauser}, M. and {Haywood}, M. and {Helmi}, A. and {Hidalgo}, S.~L. and {Hilger}, T. and {H{\l}adczuk}, N. and {Hobbs}, D. and {Holland}, G. and {Huckle}, H.~E. and {Jasniewicz}, G. and {Jonker}, P.~G. and {Juaristi Campillo}, J. and {Julbe}, F. and {Karbevska}, L. and {Kervella}, P. and {Khanna}, S. and {Kochoska}, A. and {Kontizas}, M. and {Kordopatis}, G. and {Korn}, A.~J. and {Kostrzewa-Rutkowska}, Z. and {Kruszy{\'n}ska}, K. and {Lambert}, S. and {Lanza}, A.~F. and {Lasne}, Y. and {Le Campion}, J. -F. and {Le Fustec}, Y. and {Lebreton}, Y. and {Lebzelter}, T. and {Leccia}, S. and {Leclerc}, N. and {Lecoeur-Taibi}, I. and {Liao}, S. and {Licata}, E. and {Lindstr{\o}m}, E.~P. and {Lister}, T.~A. and {Livanou}, E. and {Lobel}, A. and {Madrero Pardo}, P. and {Managau}, S. and {Mann}, R.~G. and {Marchant}, J.~M. and {Marconi}, M. and {Marcos Santos}, M.~M.~S. and {Marinoni}, S. and {Marocco}, F. and {Marshall}, D.~J. and {Martin Polo}, L. and {Mart{\'\i}n-Fleitas}, J.~M. and {Masip}, A. and {Massari}, D. and {Mastrobuono-Battisti}, A. and {Mazeh}, T. and {McMillan}, P.~J. and {Messina}, S. and {Michalik}, D. and {Millar}, N.~R. and {Mints}, A. and {Molina}, D. and {Molinaro}, R. and {Moln{\'a}r}, L. and {Montegriffo}, P. and {Mor}, R. and {Morbidelli}, R. and {Morel}, T. and {Morris}, D. and {Mulone}, A.~F. and {Munoz}, D. and {Muraveva}, T. and {Murphy}, C.~P. and {Musella}, I. and {Noval}, L. and {Ord{\'e}novic}, C. and {Orr{\`u}}, G. and {Osinde}, J. and {Pagani}, C. and {Pagano}, I. and {Palaversa}, L. and {Palicio}, P.~A. and {Panahi}, A. and {Pawlak}, M. and {Pe{\~n}alosa Esteller}, X. and {Penttil{\"a}}, A. and {Piersimoni}, A.~M. and {Pineau}, F. -X. and {Plachy}, E. and {Plum}, G. and {Poggio}, E. and {Poretti}, E. and {Poujoulet}, E. and {Pr{\v{s}}a}, A. and {Pulone}, L. and {Racero}, E. and {Ragaini}, S. and {Rainer}, M. and {Raiteri}, C.~M. and {Rambaux}, N. and {Ramos}, P. and {Ramos-Lerate}, M. and {Re Fiorentin}, P. and {Regibo}, S. and {Reyl{\'e}}, C. and {Ripepi}, V. and {Riva}, A. and {Rixon}, G. and {Robichon}, N. and {Robin}, C. and {Roelens}, M. and {Rohrbasser}, L. and {Romero-G{\'o}mez}, M. and {Rowell}, N. and {Royer}, F. and {Rybicki}, K.~A. and {Sadowski}, G. and {Sagrist{\`a} Sell{\'e}s}, A. and {Sahlmann}, J. and {Salgado}, J. and {Salguero}, E. and {Samaras}, N. and {Sanchez Gimenez}, V. and {Sanna}, N. and {Santove{\~n}a}, R. and {Sarasso}, M. and {Schultheis}, M. and {Sciacca}, E. and {Segol}, M. and {Segovia}, J.~C. and {S{\'e}gransan}, D. and {Semeux}, D. and {Shahaf}, S. and {Siddiqui}, H.~I. and {Siebert}, A. and {Siltala}, L. and {Slezak}, E. and {Smart}, R.~L. and {Solano}, E. and {Solitro}, F. and {Souami}, D. and {Souchay}, J. and {Spagna}, A. and {Spoto}, F. and {Steele}, I.~A. and {Steidelm{\"u}ller}, H. and {Stephenson}, C.~A. and {S{\"u}veges}, M. and {Szabados}, L. and {Szegedi-Elek}, E. and {Taris}, F. and {Tauran}, G. and {Taylor}, M.~B. and {Teixeira}, R. and {Thuillot}, W. and {Tonello}, N. and {Torra}, F. and {Torra}, J. and {Turon}, C. and {Unger}, N. and {Vaillant}, M. and {van Dillen}, E. and {Vanel}, O. and {Vecchiato}, A. and {Viala}, Y. and {Vicente}, D. and {Voutsinas}, S. and {Weiler}, M. and {Wevers}, T. and {Wyrzykowski}, {\L}. and {Yoldas}, A. and {Yvard}, P. and {Zhao}, H. and {Zorec}, J. and {Zucker}, S. and {Zurbach}, C. and {Zwitter}, T.},
        title = "{Gaia Early Data Release 3. Summary of the contents and survey properties}",
      journal = {\aap},
         year = 2021,
        month = may,
       volume = {649},
          eid = {A1},
        pages = {A1},
          doi = {10.1051/0004-6361/202039657},
archivePrefix = {arXiv},
       eprint = {2012.01533},
 primaryClass = {astro-ph.GA},
       adsurl = {https://ui.adsabs.harvard.edu/abs/2021A&A...649A...1G}
}

@ARTICLE{Blanco_Cuaresma_2014,
       author = {{Blanco-Cuaresma}, S. and {Soubiran}, C. and {Heiter}, U. and {Jofr{\'e}}, P.},
        title = "{Determining stellar atmospheric parameters and chemical abundances of FGK stars with iSpec}",
      journal = {\aap},
         year = 2014,
        month = sep,
       volume = {569},
          eid = {A111},
        pages = {A111},
          doi = {10.1051/0004-6361/201423945},
archivePrefix = {arXiv},
       eprint = {1407.2608},
 primaryClass = {astro-ph.IM},
       adsurl = {https://ui.adsabs.harvard.edu/abs/2014A&A...569A.111B}
}

@ARTICLE{Blanco_Cuaresma_2019,
       author = {{Blanco-Cuaresma}, Sergi},
        title = "{Modern stellar spectroscopy caveats}",
      journal = {\mnras},
         year = 2019,
        month = jun,
       volume = {486},
       number = {2},
        pages = {2075-2101},
          doi = {10.1093/mnras/stz549},
archivePrefix = {arXiv},
       eprint = {1902.09558},
 primaryClass = {astro-ph.SR},
       adsurl = {https://ui.adsabs.harvard.edu/abs/2019MNRAS.486.2075B}
}

@article{Ryabchikova_2015,
doi = {10.1088/0031-8949/90/5/054005},
url = {https://dx.doi.org/10.1088/0031-8949/90/5/054005},
year = {2015},
month = {apr},
publisher = {IOP Publishing},
volume = {90},
number = {5},
pages = {054005},
author = {T Ryabchikova and N Piskunov and R L Kurucz and H C Stempels and U Heiter and Yu Pakhomov and P S Barklem},
title = {A major upgrade of the VALD database},
journal = {Phys. Scr}
}

@ARTICLE{Asplund_2005,
       author = {{Asplund}, M. and {Grevesse}, N. and {Sauval}, A.~J.},
       title = "{The Solar Chemical Composition}",
       journal = {ASP Conf. Ser.},
       year = 2005,
       volume = {336},
        month = sep,
        pages = {25},
       adsurl = {https://ui.adsabs.harvard.edu/abs/2005ASPC..336...25A}
}

@INPROCEEDINGS{castelli_2003,
       author = {{Castelli}, F. and {Kurucz}, R.~L.},
        title = "{New Grids of ATLAS9 Model Atmospheres}",
    booktitle = {Modelling of Stellar Atmospheres},
         year = 2003,
       editor = {{Piskunov}, N. and {Weiss}, W.~W. and {Gray}, D.~F.},
       volume = {210},
        month = jan,
        pages = {A20},
          doi = {10.48550/arXiv.astro-ph/0405087},
archivePrefix = {arXiv},
       eprint = {astro-ph/0405087},
 primaryClass = {astro-ph},
       adsurl = {https://ui.adsabs.harvard.edu/abs/2003IAUS..210P.A20C}
}

@ARTICLE{Kurucz_1993,
       author = {{Kurucz}, Robert},
        title = "{SYNTHE Spectrum Synthesis Programs and Line Data.}",
      journal = {Robert Kurucz CD-ROM},
         year = 1993,
        month = jan,
       volume = {18},
       adsurl = {https://ui.adsabs.harvard.edu/abs/1993KurCD..18.....K}
}

@ARTICLE{Minniti_2024,
       author = {{Minniti}, Dante and {Matsunaga}, Noriyuki and {Fern{\'a}ndez-Trincado}, Jos{\'e} G. and {Otsubo}, Shogo and {Sarugaku}, Yuki and {Takeuchi}, Tomomi and {Katoh}, Haruki and {Hamano}, Satoshi and {Ikeda}, Yuji and {Kawakita}, Hideyo and {Lucas}, Philip W. and {Smith}, Leigh C. and {Petralia}, Ilaria and {Rita Garro}, Elisa and {Saito}, Roberto K. and {Alonso-Garc{\'\i}a}, Javier and {G{\'o}mez}, Mat{\'\i}as and {Gabriela Navarro}, Mar{\'\i}a},
        title = "{The globular cluster VVV CL002 falling down to the hazardous Galactic centre}",
      journal = {\aap},
         year = 2024,
        month = mar,
       volume = {683},
          eid = {A150},
        pages = {A150},
          doi = {10.1051/0004-6361/202348100},
archivePrefix = {arXiv},
       eprint = {2312.16028},
 primaryClass = {astro-ph.GA},
       adsurl = {https://ui.adsabs.harvard.edu/abs/2024A&A...683A.150M}
}

@ARTICLE{Sameshima_2018,
       author = {{Sameshima}, Hiroaki and {Matsunaga}, Noriyuki and {Kobayashi}, Naoto and {Kawakita}, Hideyo and {Hamano}, Satoshi and {Ikeda}, Yuji and {Kondo}, Sohei and {Fukue}, Kei and {Taniguchi}, Daisuke and {Mizumoto}, Misaki and {Arai}, Akira and {Otsubo}, Shogo and {Takenaka}, Keiichi and {Watase}, Ayaka and {Asano}, Akira and {Yasui}, Chikako and {Izumi}, Natsuko and {Yoshikawa}, Tomohiro},
        title = "{Correction of Near-infrared High-resolution Spectra for Telluric Absorption at 0.90-1.35 {\ensuremath{\mu}}m}",
      journal = {\pasp},
         year = 2018,
        month = jul,
       volume = {130},
       number = {989},
        pages = {074502},
          doi = {10.1088/1538-3873/aac1b4},
archivePrefix = {arXiv},
       eprint = {1805.00495},
 primaryClass = {astro-ph.IM},
       adsurl = {https://ui.adsabs.harvard.edu/abs/2018PASP..130g4502S}
}

@ARTICLE{Kondo_2019,
       author = {{Kondo}, Sohei and {Fukue}, Kei and {Matsunaga}, Noriyuki and {Ikeda}, Yuji and {Taniguchi}, Daisuke and {Kobayashi}, Naoto and {Sameshima}, Hiroaki and {Hamano}, Satoshi and {Arai}, Akira and {Kawakita}, Hideyo and {Yasui}, Chikako and {Izumi}, Natsuko and {Mizumoto}, Misaki and {Otsubo}, Shogo and {Takenaka}, Keiichi and {Watase}, Ayaka and {Asano}, Akira and {Yoshikawa}, Tomohiro and {Tsujimoto}, Takuji},
        title = "{Fe I Lines in 0.91-1.33 {\ensuremath{\mu}}m Spectra of Red Giants for Measuring the Microturbulence and Metallicities}",
      journal = {\apj},
         year = 2019,
        month = apr,
       volume = {875},
       number = {2},
          eid = {129},
        pages = {129},
          doi = {10.3847/1538-4357/ab0ec4},
archivePrefix = {arXiv},
       eprint = {1903.02241},
 primaryClass = {astro-ph.SR},
       adsurl = {https://ui.adsabs.harvard.edu/abs/2019ApJ...875..129K}
}

@ARTICLE{Froebrich_2007,
       author = {{Froebrich}, D. and {Scholz}, A. and {Raftery}, C.~L.},
        title = "{A systematic survey for infrared star clusters with |b| <20{\textdegree} using 2MASS}",
      journal = {\mnras},
         year = 2007,
        month = jan,
       volume = {374},
       number = {2},
        pages = {399-408},
          doi = {10.1111/j.1365-2966.2006.11148.x},
archivePrefix = {arXiv},
       eprint = {astro-ph/0610146},
 primaryClass = {astro-ph},
       adsurl = {https://ui.adsabs.harvard.edu/abs/2007MNRAS.374..399F}
}

@article{He_2023,
doi = {10.3847/1538-4365/acd6fa},
url = {https://dx.doi.org/10.3847/1538-4365/acd6fa},
year = {2023},
month = {jul},
publisher = {The American Astronomical Society},
volume = {267},
number = {2},
pages = {34},
author = {Zhihong He and Yangping Luo and Kun Wang and Anbing Ren and Liming Peng and Qian Cui and Xiaochen Liu and Qingquan Jiang},
title = {Survey for Distant Stellar Aggregates in the Galactic Disk: Detecting 2000 Star Clusters and Candidates, along with the Dwarf Galaxy IC 10},
journal = {ApJS}
}

@ARTICLE{Sagar_1997,
       author = {{Sagar}, Ram},
        title = "{The Ages of the Galactic Globular Clusters}",
      journal = {Journal of Astrophysics and Astronomy},
         year = 1997,
        month = dec,
       volume = {18},
       number = {4},
        pages = {295-301},
          doi = {10.1007/BF02709318},
       adsurl = {https://ui.adsabs.harvard.edu/abs/1997JApA...18..295S}
}

@article{West_2004,
title = "Reconstructing galaxy histories from globular clusters",
author = "West, {Michael J.} and Patrick C{\^o}t{\'e} and Marzke, {Ronald O.} and Andr{\'e}s Jord{\'a}n",
year = "2004",
month = jan,
day = "1",
doi = "10.1038/nature02235",
language = "English",
volume = "427",
pages = "31--35",
journal = "Nature",
issn = "0028-0836",
publisher = "Nature Publishing Group",
number = "6969",
}

@article{MINNITI_2010,
title = {VISTA Variables in the Via Lactea (VVV): The public ESO near-IR variability survey of the Milky Way},
journal = {New Astronomy},
volume = {15},
number = {5},
pages = {433-443},
year = {2010},
issn = {1384-1076},
doi = {https://doi.org/10.1016/j.newast.2009.12.002},
url = {https://www.sciencedirect.com/science/article/pii/S1384107609001717},
author = {D. Minniti and P.W. Lucas and J.P. Emerson and R.K. Saito and M. Hempel and P. Pietrukowicz and A.V. Ahumada and M.V. Alonso and J. Alonso-Garcia and J.I. Arias and R.M. Bandyopadhyay and R.H. Barbá and B. Barbuy and L.R. Bedin and E. Bica and J. Borissova and L. Bronfman and G. Carraro and M. Catelan and J.J. Clariá and N. Cross and R. {de Grijs} and I. Dékány and J.E. Drew and C. Fariña and C. Feinstein and E. Fernández Lajús and R.C. Gamen and D. Geisler and W. Gieren and B. Goldman and O.A. Gonzalez and G. Gunthardt and S. Gurovich and N.C. Hambly and M.J. Irwin and V.D. Ivanov and A. Jordán and E. Kerins and K. Kinemuchi and R. Kurtev and M. López-Corredoira and T. Maccarone and N. Masetti and D. Merlo and M. Messineo and I.F. Mirabel and L. Monaco and L. Morelli and N. Padilla and T. Palma and M.C. Parisi and G. Pignata and M. Rejkuba and A. Roman-Lopes and S.E. Sale and M.R. Schreiber and A.C. Schröder and M. Smith and L. Sodré Jr. and M. Soto and M. Tamura and C. Tappert and M.A. Thompson and I. Toledo and M. Zoccali and G. Pietrzynski}
}

@ARTICLE{Saito_2024,
       author = {{Saito}, R.~K. and {Hempel}, M. and {Alonso-Garc{\'\i}a}, J. and {Lucas}, P.~W. and {Minniti}, D. and {Alonso}, S. and {Baravalle}, L. and {Borissova}, J. and {Caceres}, C. and {Chen{\'e}}, A.~N. and {Cross}, N.~J.~G. and {Duplancic}, F. and {Garro}, E.~R. and {G{\'o}mez}, M. and {Ivanov}, V.~D. and {Kurtev}, R. and {Luna}, A. and {Majaess}, D. and {Navarro}, M.~G. and {Pullen}, J.~B. and {Rejkuba}, M. and {Sanders}, J.~L. and {Smith}, L.~C. and {Albino}, P.~H.~C. and {Alonso}, M.~V. and {Am{\^o}res}, E.~B. and {Angeloni}, R. and {Arias}, J.~I. and {Arnaboldi}, M. and {Barbuy}, B. and {Bayo}, A. and {Beamin}, J.~C. and {Bedin}, L.~R. and {Bellini}, A. and {Benjamin}, R.~A. and {Bica}, E. and {Bonatto}, C.~J. and {Botan}, E. and {Braga}, V.~F. and {Brown}, D.~A. and {Cabral}, J.~B. and {Camargo}, D. and {Caratti o Garatti}, A. and {Carballo-Bello}, J.~A. and {Catelan}, M. and {Chavero}, C. and {Chijani}, M.~A. and {Clari{\'a}}, J.~J. and {Coldwell}, G.~V. and {Pe{\~n}a}, C. Contreras and {Ramos}, R. Contreras and {Corral-Santana}, J.~M. and {Cort{\'e}s}, C.~C. and {Cort{\'e}s-Contreras}, M. and {Cruz}, P. and {Daza-Perilla}, I.~V. and {Debattista}, V.~P. and {Dias}, B. and {Donoso}, L. and {D'Souza}, R. and {Emerson}, J.~P. and {Federle}, S. and {Fermiano}, V. and {Fernandez}, J. and {Fern{\'a}ndez-Trincado}, J.~G. and {Ferreira}, T. and {Lopes}, C.~E. Ferreira and {Firpo}, V. and {Flores-Quintana}, C. and {Fraga}, L. and {Froebrich}, D. and {Galdeano}, D. and {Gavignaud}, I. and {Geisler}, D. and {Gerhard}, O.~E. and {Gieren}, W. and {Gonzalez}, O.~A. and {Gramajo}, L.~V. and {Gran}, F. and {Granitto}, P.~M. and {Griggio}, M. and {Guo}, Z. and {Gurovich}, S. and {Hilker}, M. and {Jones}, H.~R.~A. and {Kammers}, R. and {Kuhn}, M.~A. and {Kumar}, M.~S.~N. and {Kundu}, R. and {Lares}, M. and {Libralato}, M. and {Lima}, E. and {Maccarone}, T.~J. and {Cort{\'e}s}, P. Marchant and {Martin}, E.~L. and {Masetti}, N. and {Matsunaga}, N. and {Mauro}, F. and {McDonald}, I. and {Mej{\'\i}as}, A. and {Mesa}, V. and {Milla-Castro}, F.~P. and {Minniti}, J.~H. and {Bidin}, C. Moni and {Montenegro}, K. and {Morris}, C. and {Motta}, V. and {Navarete}, F. and {Molina}, C. Navarro and {Nikzat}, F. and {Castell{\'o}n}, J.~L. Nilo and {Obasi}, C. and {Ortigoza-Urdaneta}, M. and {Palma}, T. and {Parisi}, C. and {Ram{\'\i}rez}, K. Pena and {Pereyra}, L. and {Perez}, N. and {Petralia}, I. and {Pichel}, A. and {Pignata}, G. and {Alegr{\'\i}a}, S. Ram{\'\i}rez and {Rojas}, A.~F. and {Rojas}, D. and {Roman-Lopes}, A. and {Rovero}, A.~C. and {Saroon}, S. and {Schmidt}, E.~O. and {Schr{\"o}der}, A.~C. and {Schultheis}, M. and {Sgr{\'o}}, M.~A. and {Solano}, E. and {Soto}, M. and {Stecklum}, B. and {Steeghs}, D. and {Tamura}, M. and {Tissera}, P. and {Valcarce}, A.~A.~R. and {Valotto}, C.~A. and {Vasquez}, S. and {Villalon}, C. and {Villanova}, S. and {C{\'a}diz}, F. Vivanco and {Bacigalupo}, R. Zelada and {Zijlstra}, A. and {Zoccali}, M.},
        title = "{The VISTA Variables in the V{\'\i}a L{\'a}ctea extended (VVVX) ESO public survey: Completion of the observations and legacy}",
      journal = {\aap},
         year = 2024,
        month = sep,
       volume = {689},
          eid = {A148},
        pages = {A148},
          doi = {10.1051/0004-6361/202450584},
archivePrefix = {arXiv},
       eprint = {2406.16646},
 primaryClass = {astro-ph.GA},
       adsurl = {https://ui.adsabs.harvard.edu/abs/2024A&A...689A.148S}
}

@ARTICLE{Sanders_2019,
       author = {{Sanders}, Jason L. and {Smith}, Leigh and {Evans}, N. Wyn},
        title = "{The pattern speed of the Milky Way bar from transverse velocities}",
      journal = {\mnras},
         year = 2019,
        month = oct,
       volume = {488},
       number = {4},
        pages = {4552-4564},
          doi = {10.1093/mnras/stz1827},
archivePrefix = {arXiv},
       eprint = {1903.02009},
 primaryClass = {astro-ph.GA},
       adsurl = {https://ui.adsabs.harvard.edu/abs/2019MNRAS.488.4552S}
}

@ARTICLE{Smirnov_2024,
       author = {{Smirnov}, Anton A. and {Bajkova}, Anisa T. and {Bobylev}, Vadim V.},
        title = "{Globular clusters and bar: captured or not captured?}",
      journal = {\mnras},
         year = 2024,
        month = feb,
       volume = {528},
       number = {2},
        pages = {1422-1437},
          doi = {10.1093/mnras/stae029},
archivePrefix = {arXiv},
       eprint = {2310.18172},
 primaryClass = {astro-ph.GA},
       adsurl = {https://ui.adsabs.harvard.edu/abs/2024MNRAS.528.1422S}
}

@ARTICLE{Gaia_DR3,
       author = {{Gaia Collaboration} and {Vallenari}, A. and {Brown}, A.~G.~A. and {Prusti}, T. and {de Bruijne}, J.~H.~J. and {Arenou}, F. and {Babusiaux}, C. and {Biermann}, M. and {Creevey}, O.~L. and {Ducourant}, C. and {Evans}, D.~W. and {Eyer}, L. and {Guerra}, R. and {Hutton}, A. and {Jordi}, C. and {Klioner}, S.~A. and {Lammers}, U.~L. and {Lindegren}, L. and {Luri}, X. and {Mignard}, F. and {Panem}, C. and {Pourbaix}, D. and {Randich}, S. and {Sartoretti}, P. and {Soubiran}, C. and {Tanga}, P. and {Walton}, N.~A. and {Bailer-Jones}, C.~A.~L. and {Bastian}, U. and {Drimmel}, R. and {Jansen}, F. and {Katz}, D. and {Lattanzi}, M.~G. and {van Leeuwen}, F. and {Bakker}, J. and {Cacciari}, C. and {Casta{\~n}eda}, J. and {De Angeli}, F. and {Fabricius}, C. and {Fouesneau}, M. and {Fr{\'e}mat}, Y. and {Galluccio}, L. and {Guerrier}, A. and {Heiter}, U. and {Masana}, E. and {Messineo}, R. and {Mowlavi}, N. and {Nicolas}, C. and {Nienartowicz}, K. and {Pailler}, F. and {Panuzzo}, P. and {Riclet}, F. and {Roux}, W. and {Seabroke}, G.~M. and {Sordo}, R. and {Th{\'e}venin}, F. and {Gracia-Abril}, G. and {Portell}, J. and {Teyssier}, D. and {Altmann}, M. and {Andrae}, R. and {Audard}, M. and {Bellas-Velidis}, I. and {Benson}, K. and {Berthier}, J. and {Blomme}, R. and {Burgess}, P.~W. and {Busonero}, D. and {Busso}, G. and {C{\'a}novas}, H. and {Carry}, B. and {Cellino}, A. and {Cheek}, N. and {Clementini}, G. and {Damerdji}, Y. and {Davidson}, M. and {de Teodoro}, P. and {Nu{\~n}ez Campos}, M. and {Delchambre}, L. and {Dell'Oro}, A. and {Esquej}, P. and {Fern{\'a}ndez-Hern{\'a}ndez}, J. and {Fraile}, E. and {Garabato}, D. and {Garc{\'\i}a-Lario}, P. and {Gosset}, E. and {Haigron}, R. and {Halbwachs}, J. -L. and {Hambly}, N.~C. and {Harrison}, D.~L. and {Hern{\'a}ndez}, J. and {Hestroffer}, D. and {Hodgkin}, S.~T. and {Holl}, B. and {Jan{\ss}en}, K. and {Jevardat de Fombelle}, G. and {Jordan}, S. and {Krone-Martins}, A. and {Lanzafame}, A.~C. and {L{\"o}ffler}, W. and {Marchal}, O. and {Marrese}, P.~M. and {Moitinho}, A. and {Muinonen}, K. and {Osborne}, P. and {Pancino}, E. and {Pauwels}, T. and {Recio-Blanco}, A. and {Reyl{\'e}}, C. and {Riello}, M. and {Rimoldini}, L. and {Roegiers}, T. and {Rybizki}, J. and {Sarro}, L.~M. and {Siopis}, C. and {Smith}, M. and {Sozzetti}, A. and {Utrilla}, E. and {van Leeuwen}, M. and {Abbas}, U. and {{\'A}brah{\'a}m}, P. and {Abreu Aramburu}, A. and {Aerts}, C. and {Aguado}, J.~J. and {Ajaj}, M. and {Aldea-Montero}, F. and {Altavilla}, G. and {{\'A}lvarez}, M.~A. and {Alves}, J. and {Anders}, F. and {Anderson}, R.~I. and {Anglada Varela}, E. and {Antoja}, T. and {Baines}, D. and {Baker}, S.~G. and {Balaguer-N{\'u}{\~n}ez}, L. and {Balbinot}, E. and {Balog}, Z. and {Barache}, C. and {Barbato}, D. and {Barros}, M. and {Barstow}, M.~A. and {Bartolom{\'e}}, S. and {Bassilana}, J. -L. and {Bauchet}, N. and {Becciani}, U. and {Bellazzini}, M. and {Berihuete}, A. and {Bernet}, M. and {Bertone}, S. and {Bianchi}, L. and {Binnenfeld}, A. and {Blanco-Cuaresma}, S. and {Blazere}, A. and {Boch}, T. and {Bombrun}, A. and {Bossini}, D. and {Bouquillon}, S. and {Bragaglia}, A. and {Bramante}, L. and {Breedt}, E. and {Bressan}, A. and {Brouillet}, N. and {Brugaletta}, E. and {Bucciarelli}, B. and {Burlacu}, A. and {Butkevich}, A.~G. and {Buzzi}, R. and {Caffau}, E. and {Cancelliere}, R. and {Cantat-Gaudin}, T. and {Carballo}, R. and {Carlucci}, T. and {Carnerero}, M.~I. and {Carrasco}, J.~M. and {Casamiquela}, L. and {Castellani}, M. and {Castro-Ginard}, A. and {Chaoul}, L. and {Charlot}, P. and {Chemin}, L. and {Chiaramida}, V. and {Chiavassa}, A. and {Chornay}, N. and {Comoretto}, G. and {Contursi}, G. and {Cooper}, W.~J. and {Cornez}, T. and {Cowell}, S. and {Crifo}, F. and {Cropper}, M. and {Crosta}, M. and {Crowley}, C. and {Dafonte}, C. and {Dapergolas}, A. and {David}, M. and {David}, P. and {de Laverny}, P. and {De Luise}, F. and {De March}, R. and {De Ridder}, J. and {de Souza}, R. and {de Torres}, A. and {del Peloso}, E.~F. and {del Pozo}, E. and {Delbo}, M. and {Delgado}, A. and {Delisle}, J. -B. and {Demouchy}, C. and {Dharmawardena}, T.~E. and {Di Matteo}, P. and {Diakite}, S. and {Diener}, C. and {Distefano}, E. and {Dolding}, C. and {Edvardsson}, B. and {Enke}, H. and {Fabre}, C. and {Fabrizio}, M. and {Faigler}, S. and {Fedorets}, G. and {Fernique}, P. and {Fienga}, A. and {Figueras}, F. and {Fournier}, Y. and {Fouron}, C. and {Fragkoudi}, F. and {Gai}, M. and {Garcia-Gutierrez}, A. and {Garcia-Reinaldos}, M. and {Garc{\'\i}a-Torres}, M. and {Garofalo}, A. and {Gavel}, A. and {Gavras}, P. and {Gerlach}, E. and {Geyer}, R. and {Giacobbe}, P. and {Gilmore}, G. and {Girona}, S. and {Giuffrida}, G. and {Gomel}, R. and {Gomez}, A. and {Gonz{\'a}lez-N{\'u}{\~n}ez}, J. and {Gonz{\'a}lez-Santamar{\'\i}a}, I. and {Gonz{\'a}lez-Vidal}, J.~J. and {Granvik}, M. and {Guillout}, P. and {Guiraud}, J. and {Guti{\'e}rrez-S{\'a}nchez}, R. and {Guy}, L.~P. and {Hatzidimitriou}, D. and {Hauser}, M. and {Haywood}, M. and {Helmer}, A. and {Helmi}, A. and {Sarmiento}, M.~H. and {Hidalgo}, S.~L. and {Hilger}, T. and {H{\l}adczuk}, N. and {Hobbs}, D. and {Holland}, G. and {Huckle}, H.~E. and {Jardine}, K. and {Jasniewicz}, G. and {Jean-Antoine Piccolo}, A. and {Jim{\'e}nez-Arranz}, {\'O}. and {Jorissen}, A. and {Juaristi Campillo}, J. and {Julbe}, F. and {Karbevska}, L. and {Kervella}, P. and {Khanna}, S. and {Kontizas}, M. and {Kordopatis}, G. and {Korn}, A.~J. and {K{\'o}sp{\'a}l}, {\'A}. and {Kostrzewa-Rutkowska}, Z. and {Kruszy{\'n}ska}, K. and {Kun}, M. and {Laizeau}, P. and {Lambert}, S. and {Lanza}, A.~F. and {Lasne}, Y. and {Le Campion}, J. -F. and {Lebreton}, Y. and {Lebzelter}, T. and {Leccia}, S. and {Leclerc}, N. and {Lecoeur-Taibi}, I. and {Liao}, S. and {Licata}, E.~L. and {Lindstr{\o}m}, H.~E.~P. and {Lister}, T.~A. and {Livanou}, E. and {Lobel}, A. and {Lorca}, A. and {Loup}, C. and {Madrero Pardo}, P. and {Magdaleno Romeo}, A. and {Managau}, S. and {Mann}, R.~G. and {Manteiga}, M. and {Marchant}, J.~M. and {Marconi}, M. and {Marcos}, J. and {Marcos Santos}, M.~M.~S. and {Mar{\'\i}n Pina}, D. and {Marinoni}, S. and {Marocco}, F. and {Marshall}, D.~J. and {Martin Polo}, L. and {Mart{\'\i}n-Fleitas}, J.~M. and {Marton}, G. and {Mary}, N. and {Masip}, A. and {Massari}, D. and {Mastrobuono-Battisti}, A. and {Mazeh}, T. and {McMillan}, P.~J. and {Messina}, S. and {Michalik}, D. and {Millar}, N.~R. and {Mints}, A. and {Molina}, D. and {Molinaro}, R. and {Moln{\'a}r}, L. and {Monari}, G. and {Mongui{\'o}}, M. and {Montegriffo}, P. and {Montero}, A. and {Mor}, R. and {Mora}, A. and {Morbidelli}, R. and {Morel}, T. and {Morris}, D. and {Muraveva}, T. and {Murphy}, C.~P. and {Musella}, I. and {Nagy}, Z. and {Noval}, L. and {Oca{\~n}a}, F. and {Ogden}, A. and {Ordenovic}, C. and {Osinde}, J.~O. and {Pagani}, C. and {Pagano}, I. and {Palaversa}, L. and {Palicio}, P.~A. and {Pallas-Quintela}, L. and {Panahi}, A. and {Payne-Wardenaar}, S. and {Pe{\~n}alosa Esteller}, X. and {Penttil{\"a}}, A. and {Pichon}, B. and {Piersimoni}, A.~M. and {Pineau}, F. -X. and {Plachy}, E. and {Plum}, G. and {Poggio}, E. and {Pr{\v{s}}a}, A. and {Pulone}, L. and {Racero}, E. and {Ragaini}, S. and {Rainer}, M. and {Raiteri}, C.~M. and {Rambaux}, N. and {Ramos}, P. and {Ramos-Lerate}, M. and {Re Fiorentin}, P. and {Regibo}, S. and {Richards}, P.~J. and {Rios Diaz}, C. and {Ripepi}, V. and {Riva}, A. and {Rix}, H. -W. and {Rixon}, G. and {Robichon}, N. and {Robin}, A.~C. and {Robin}, C. and {Roelens}, M. and {Rogues}, H.~R.~O. and {Rohrbasser}, L. and {Romero-G{\'o}mez}, M. and {Rowell}, N. and {Royer}, F. and {Ruz Mieres}, D. and {Rybicki}, K.~A. and {Sadowski}, G. and {S{\'a}ez N{\'u}{\~n}ez}, A. and {Sagrist{\`a} Sell{\'e}s}, A. and {Sahlmann}, J. and {Salguero}, E. and {Samaras}, N. and {Sanchez Gimenez}, V. and {Sanna}, N. and {Santove{\~n}a}, R. and {Sarasso}, M. and {Schultheis}, M. and {Sciacca}, E. and {Segol}, M. and {Segovia}, J.~C. and {S{\'e}gransan}, D. and {Semeux}, D. and {Shahaf}, S. and {Siddiqui}, H.~I. and {Siebert}, A. and {Siltala}, L. and {Silvelo}, A. and {Slezak}, E. and {Slezak}, I. and {Smart}, R.~L. and {Snaith}, O.~N. and {Solano}, E. and {Solitro}, F. and {Souami}, D. and {Souchay}, J. and {Spagna}, A. and {Spina}, L. and {Spoto}, F. and {Steele}, I.~A. and {Steidelm{\"u}ller}, H. and {Stephenson}, C.~A. and {S{\"u}veges}, M. and {Surdej}, J. and {Szabados}, L. and {Szegedi-Elek}, E. and {Taris}, F. and {Taylor}, M.~B. and {Teixeira}, R. and {Tolomei}, L. and {Tonello}, N. and {Torra}, F. and {Torra}, J. and {Torralba Elipe}, G. and {Trabucchi}, M. and {Tsounis}, A.~T. and {Turon}, C. and {Ulla}, A. and {Unger}, N. and {Vaillant}, M.~V. and {van Dillen}, E. and {van Reeven}, W. and {Vanel}, O. and {Vecchiato}, A. and {Viala}, Y. and {Vicente}, D. and {Voutsinas}, S. and {Weiler}, M. and {Wevers}, T. and {Wyrzykowski}, {\L}. and {Yoldas}, A. and {Yvard}, P. and {Zhao}, H. and {Zorec}, J. and {Zucker}, S. and {Zwitter}, T.},
        title = "{Gaia Data Release 3. Summary of the content and survey properties}",
      journal = {\aap},
         year = 2023,
        month = jun,
       volume = {674},
          eid = {A1},
        pages = {A1},
          doi = {10.1051/0004-6361/202243940},
archivePrefix = {arXiv},
       eprint = {2208.00211},
 primaryClass = {astro-ph.GA},
       adsurl = {https://ui.adsabs.harvard.edu/abs/2023A&A...674A...1G}
}

@ARTICLE{Zucker_2003,
       author = {{Zucker}, S.},
        title = "{Cross-correlation and maximum-likelihood analysis: a new approach to combining cross-correlation functions}",
      journal = {\mnras},
         year = 2003,
        month = jul,
       volume = {342},
       number = {4},
        pages = {1291-1298},
          doi = {10.1046/j.1365-8711.2003.06633.x},
archivePrefix = {arXiv},
       eprint = {astro-ph/0303426},
 primaryClass = {astro-ph},
       adsurl = {https://ui.adsabs.harvard.edu/abs/2003MNRAS.342.1291Z}
}

@ARTICLE{Garro_2022_Garro2,
       author = {{Garro}, E.~R. and {Minniti}, D. and {G{\'o}mez}, M. and {Fern{\'a}ndez-Trincado}, J.~G. and {Alonso-Garc{\'\i}a}, J. and {Hempel}, M. and {Zelada Bacigalupo}, R.},
        title = "{A new low-luminosity globular cluster discovered in the Milky Way with the VVVX survey}",
      journal = {\aap},
         year = 2022,
        month = jun,
       volume = {662},
          eid = {A95},
        pages = {A95},
          doi = {10.1051/0004-6361/202243342},
archivePrefix = {arXiv},
       eprint = {2205.03444},
 primaryClass = {astro-ph.GA},
       adsurl = {https://ui.adsabs.harvard.edu/abs/2022A&A...662A..95G}
}

@ARTICLE{Bonatto_2007,
       author = {{Bonatto}, C. and {Bica}, E. and {Ortolani}, S. and {Barbuy}, B.},
        title = "{FSR1767 - a new globular cluster in the Galaxy}",
      journal = {\mnras},
         year = 2007,
        month = oct,
       volume = {381},
       number = {1},
        pages = {L45-L49},
          doi = {10.1111/j.1745-3933.2007.00363.x},
archivePrefix = {arXiv},
       eprint = {0708.0501},
 primaryClass = {astro-ph},
       adsurl = {https://ui.adsabs.harvard.edu/abs/2007MNRAS.381L..45B}
}

@article{Bonatto_2009,
    author = {Bonatto, C. and Bica, E. and Ortolani, S. and Barbuy, B.},
    title = {Further probing the nature of FSR 1767},
    journal = {Monthly Notices of the Royal Astronomical Society},
    volume = {397},
    number = {2},
    pages = {1032-1040},
    year = {2009},
    month = {07},
    issn = {0035-8711},
    doi = {10.1111/j.1365-2966.2009.15020.x},
    url = {https://doi.org/10.1111/j.1365-2966.2009.15020.x},
    eprint = {https://academic.oup.com/mnras/article-pdf/397/2/1032/2945173/mnras0397-1032.pdf},
}

@ARTICLE{Buckner_&_Froebrich_2013,
       author = {{Buckner}, Anne S.~M. and {Froebrich}, Dirk},
        title = "{Properties of star clusters - I. Automatic distance and extinction estimates}",
      journal = {\mnras},
         year = 2013,
        month = dec,
       volume = {436},
       number = {2},
        pages = {1465-1478},
          doi = {10.1093/mnras/stt1665},
archivePrefix = {arXiv},
       eprint = {1309.0708},
 primaryClass = {astro-ph.GA},
       adsurl = {https://ui.adsabs.harvard.edu/abs/2013MNRAS.436.1465B}
}

@ARTICLE{Garro_2022_FSR1767,
       author = {{Garro}, E.~R. and {Minniti}, D. and {G{\'o}mez}, M. and {Alonso-Garc{\'\i}a}, J. and {Ripepi}, V. and {Fern{\'a}ndez-Trincado}, J.~G. and {Vivanco C{\'a}diz}, F.},
        title = "{Inspection of 19 globular cluster candidates in the Galactic bulge with the VVV survey}",
      journal = {\aap},
         year = 2022,
        month = feb,
       volume = {658},
          eid = {A120},
        pages = {A120},
          doi = {10.1051/0004-6361/202141819},
archivePrefix = {arXiv},
       eprint = {2111.08317},
 primaryClass = {astro-ph.GA},
       adsurl = {https://ui.adsabs.harvard.edu/abs/2022A&A...658A.120G}
}

@ARTICLE{Mercer_2005,
       author = {{Mercer}, E.~P. and {Clemens}, D.~P. and {Meade}, M.~R. and {Babler}, B.~L. and {Indebetouw}, R. and {Whitney}, B.~A. and {Watson}, C. and {Wolfire}, M.~G. and {Wolff}, M.~J. and {Bania}, T.~M. and {Benjamin}, R.~A. and {Cohen}, M. and {Dickey}, J.~M. and {Jackson}, J.~M. and {Kobulnicky}, H.~A. and {Mathis}, J.~S. and {Stauffer}, J.~R. and {Stolovy}, S.~R. and {Uzpen}, B. and {Churchwell}, E.~B.},
        title = "{New Star Clusters Discovered in the GLIMPSE Survey}",
      journal = {\apj},
         year = 2005,
        month = dec,
       volume = {635},
       number = {1},
        pages = {560-569},
          doi = {10.1086/497260},
       adsurl = {https://ui.adsabs.harvard.edu/abs/2005ApJ...635..560M}
}

@ARTICLE{Robin2012,
       author = {{Robin}, A.~C. and {Marshall}, D.~J. and {Schultheis}, M. and {Reyl{\'e}}, C.},
        title = "{Stellar populations in the Milky Way bulge region: towards solving the Galactic bulge and bar shapes using 2MASS data}",
      journal = {\aap},
         year = 2012,
        month = feb,
       volume = {538},
          eid = {A106},
        pages = {A106},
          doi = {10.1051/0004-6361/201116512},
archivePrefix = {arXiv},
       eprint = {1111.5744},
 primaryClass = {astro-ph.GA},
       adsurl = {https://ui.adsabs.harvard.edu/abs/2012A&A...538A.106R}
}

@ARTICLE{Fernandez-Trincado2019,
       author = {{Fern{\'a}ndez-Trincado}, Jos{\'e} G. and {Beers}, Timothy C. and {Tang}, Baitian and {Moreno}, Edmundo and {P{\'e}rez-Villegas}, Angeles and {Ortigoza-Urdaneta}, Mario},
        title = "{Chemodynamics of newly identified giants with a globular cluster like abundance patterns in the bulge, disc, and halo of the Milky Way}",
      journal = {\mnras},
         year = 2019,
        month = sep,
       volume = {488},
       number = {2},
        pages = {2864-2880},
          doi = {10.1093/mnras/stz1848},
archivePrefix = {arXiv},
       eprint = {1904.05369},
 primaryClass = {astro-ph.GA},
       adsurl = {https://ui.adsabs.harvard.edu/abs/2019MNRAS.488.2864F}
}

@ARTICLE{Tang2018,
       author = {{Tang}, Baitian and {Fern{\'a}ndez-Trincado}, J.~G. and {Geisler}, Doug and {Zamora}, Olga and {M{\'e}sz{\'a}ros}, Szabolcs and {Masseron}, Thomas and {Cohen}, Roger E. and {Garc{\'\i}a-Hern{\'a}ndez}, D.~A. and {Dell'Agli}, Flavia and {Beers}, Timothy C. and {Schiavon}, Ricardo P. and {Sohn}, Sangmo Tony and {Hasselquist}, Sten and {Robin}, Annie C. and {Shetrone}, Matthew and {Majewski}, Steven R. and {Villanova}, Sandro and {Schiappacasse Ulloa}, Jose and {Lane}, Richard R. and {Minnti}, Dante and {Roman-Lopes}, Alexandre and {Almeida}, Andres and {Moreno}, E.},
        title = "{The Metal-poor non-Sagittarius (?) Globular Cluster NGC 5053: Orbit and Mg, Al, and Si Abundances}",
      journal = {\apj},
         year = 2018,
        month = mar,
       volume = {855},
       number = {1},
          eid = {38},
        pages = {38},
          doi = {10.3847/1538-4357/aaaaea},
archivePrefix = {arXiv},
       eprint = {1801.08265},
 primaryClass = {astro-ph.SR},
       adsurl = {https://ui.adsabs.harvard.edu/abs/2018ApJ...855...38T}
}

@ARTICLE{Ralph2010,
       author = {{Sch{\"o}nrich}, Ralph and {Binney}, James and {Dehnen}, Walter},
        title = "{Local kinematics and the local standard of rest}",
      journal = {\mnras},
         year = 2010,
        month = apr,
       volume = {403},
       number = {4},
        pages = {1829-1833},
          doi = {10.1111/j.1365-2966.2010.16253.x},
archivePrefix = {arXiv},
       eprint = {0912.3693},
 primaryClass = {astro-ph.GA},
       adsurl = {https://ui.adsabs.harvard.edu/abs/2010MNRAS.403.1829S}
}

@ARTICLE{GRAVITY2021,
       author = {{GRAVITY Collaboration} and {Abuter}, R. and {Amorim}, A. and {Baub{\"o}ck}, M. and {Berger}, J.~P. and {Bonnet}, H. and {Brandner}, W. and {Cl{\'e}net}, Y. and {Davies}, R. and {de Zeeuw}, P.~T. and {Dexter}, J. and {Dallilar}, Y. and {Drescher}, A. and {Eckart}, A. and {Eisenhauer}, F. and {F{\"o}rster Schreiber}, N.~M. and {Garcia}, P. and {Gao}, F. and {Gendron}, E. and {Genzel}, R. and {Gillessen}, S. and {Habibi}, M. and {Haubois}, X. and {Hei{\ss}el}, G. and {Henning}, T. and {Hippler}, S. and {Horrobin}, M. and {Jim{\'e}nez-Rosales}, A. and {Jochum}, L. and {Jocou}, L. and {Kaufer}, A. and {Kervella}, P. and {Lacour}, S. and {Lapeyr{\`e}re}, V. and {Le Bouquin}, J. -B. and {L{\'e}na}, P. and {Lutz}, D. and {Nowak}, M. and {Ott}, T. and {Paumard}, T. and {Perraut}, K. and {Perrin}, G. and {Pfuhl}, O. and {Rabien}, S. and {Rodr{\'\i}guez-Coira}, G. and {Shangguan}, J. and {Shimizu}, T. and {Scheithauer}, S. and {Stadler}, J. and {Straub}, O. and {Straubmeier}, C. and {Sturm}, E. and {Tacconi}, L.~J. and {Vincent}, F. and {von Fellenberg}, S. and {Waisberg}, I. and {Widmann}, F. and {Wieprecht}, E. and {Wiezorrek}, E. and {Woillez}, J. and {Yazici}, S. and {Young}, A. and {Zins}, G.},
        title = "{Improved GRAVITY astrometric accuracy from modeling optical aberrations}",
      journal = {\aap},
         year = 2021,
        month = mar,
       volume = {647},
          eid = {A59},
        pages = {A59},
          doi = {10.1051/0004-6361/202040208},
archivePrefix = {arXiv},
       eprint = {2101.12098},
 primaryClass = {astro-ph.GA},
       adsurl = {https://ui.adsabs.harvard.edu/abs/2021A&A...647A..59G}
}

@ARTICLE{Eilers2019,
       author = {{Eilers}, Anna-Christina and {Hogg}, David W. and {Rix}, Hans-Walter and {Ness}, Melissa K.},
        title = "{The Circular Velocity Curve of the Milky Way from 5 to 25 kpc}",
      journal = {\apj},
         year = 2019,
        month = jan,
       volume = {871},
       number = {1},
          eid = {120},
        pages = {120},
          doi = {10.3847/1538-4357/aaf648},
archivePrefix = {arXiv},
       eprint = {1810.09466},
 primaryClass = {astro-ph.GA},
       adsurl = {https://ui.adsabs.harvard.edu/abs/2019ApJ...871..120E}
}

@ARTICLE{Perez-Villegas2018,
       author = {{P{\'e}rez-Villegas}, A. and {Rossi}, L. and {Ortolani}, S. and {Casotto}, S. and {Barbuy}, B. and {Bica}, E.},
        title = "{Orbits of Selected Globular Clusters in the Galactic Bulge}",
      journal = {\pasa},
         year = 2018,
        month = may,
       volume = {35},
          eid = {e021},
        pages = {e021},
          doi = {10.1017/pasa.2018.16},
archivePrefix = {arXiv},
       eprint = {1804.05781},
 primaryClass = {astro-ph.GA},
       adsurl = {https://ui.adsabs.harvard.edu/abs/2018PASA...35...21P}
}

@ARTICLE{2MASS,
       author = {{Skrutskie}, M.~F. and {Cutri}, R.~M. and {Stiening}, R. and {Weinberg}, M.~D. and {Schneider}, S. and {Carpenter}, J.~M. and {Beichman}, C. and {Capps}, R. and {Chester}, T. and {Elias}, J. and {Huchra}, J. and {Liebert}, J. and {Lonsdale}, C. and {Monet}, D.~G. and {Price}, S. and {Seitzer}, P. and {Jarrett}, T. and {Kirkpatrick}, J.~D. and {Gizis}, J.~E. and {Howard}, E. and {Evans}, T. and {Fowler}, J. and {Fullmer}, L. and {Hurt}, R. and {Light}, R. and {Kopan}, E.~L. and {Marsh}, K.~A. and {McCallon}, H.~L. and {Tam}, R. and {Van Dyk}, S. and {Wheelock}, S.},
        title = "{The Two Micron All Sky Survey (2MASS)}",
      journal = {\aj},
         year = 2006,
        month = feb,
       volume = {131},
       number = {2},
        pages = {1163-1183},
          doi = {10.1086/498708},
       adsurl = {https://ui.adsabs.harvard.edu/abs/2006AJ....131.1163S}
}

@ARTICLE{Soubiran+18,
       author = {{Soubiran}, C. and {Cantat-Gaudin}, T. and {Romero-G{\'o}mez}, M. and {Casamiquela}, L. and {Jordi}, C. and {Vallenari}, A. and {Antoja}, T. and {Balaguer-N{\'u}{\~n}ez}, L. and {Bossini}, D. and {Bragaglia}, A. and {Carrera}, R. and {Castro-Ginard}, A. and {Figueras}, F. and {Heiter}, U. and {Katz}, D. and {Krone-Martins}, A. and {Le Campion}, J. -F. and {Moitinho}, A. and {Sordo}, R.},
        title = "{Open cluster kinematics with Gaia DR2}",
      journal = {\aap},
         year = 2018,
        month = nov,
       volume = {619},
          eid = {A155},
        pages = {A155},
          doi = {10.1051/0004-6361/201834020},
archivePrefix = {arXiv},
       eprint = {1808.01613},
 primaryClass = {astro-ph.SR},
       adsurl = {https://ui.adsabs.harvard.edu/abs/2018A&A...619A.155S}
}

@ARTICLE{Kharchenko+13,
       author = {{Kharchenko}, N.~V. and {Piskunov}, A.~E. and {Schilbach}, E. and {R{\"o}ser}, S. and {Scholz}, R. -D.},
        title = "{Global survey of star clusters in the Milky Way. II. The catalogue of basic parameters}",
      journal = {\aap},
         year = 2013,
        month = oct,
       volume = {558},
          eid = {A53},
        pages = {A53},
          doi = {10.1051/0004-6361/201322302},
archivePrefix = {arXiv},
       eprint = {1308.5822},
 primaryClass = {astro-ph.GA},
       adsurl = {https://ui.adsabs.harvard.edu/abs/2013A&A...558A..53K}
}

@ARTICLE{Cantat-Gaudin+18,
       author = {{Cantat-Gaudin}, T. and {Jordi}, C. and {Vallenari}, A. and {Bragaglia}, A. and {Balaguer-N{\'u}{\~n}ez}, L. and {Soubiran}, C. and {Bossini}, D. and {Moitinho}, A. and {Castro-Ginard}, A. and {Krone-Martins}, A. and {Casamiquela}, L. and {Sordo}, R. and {Carrera}, R.},
        title = "{A Gaia DR2 view of the open cluster population in the Milky Way}",
      journal = {\aap},
         year = 2018,
        month = oct,
       volume = {618},
          eid = {A93},
        pages = {A93},
          doi = {10.1051/0004-6361/201833476},
archivePrefix = {arXiv},
       eprint = {1805.08726},
 primaryClass = {astro-ph.GA},
       adsurl = {https://ui.adsabs.harvard.edu/abs/2018A&A...618A..93C}
}

@ARTICLE{Vasiliev-Baumgardt+21,
       author = {{Vasiliev}, Eugene and {Baumgardt}, Holger},
        title = "{Gaia EDR3 view on galactic globular clusters}",
      journal = {\mnras},
         year = 2021,
        month = aug,
       volume = {505},
       number = {4},
        pages = {5978-6002},
          doi = {10.1093/mnras/stab1475},
archivePrefix = {arXiv},
       eprint = {2102.09568},
 primaryClass = {astro-ph.GA},
       adsurl = {https://ui.adsabs.harvard.edu/abs/2021MNRAS.505.5978V}
}

@ARTICLE{Belokurov+24,
       author = {{Belokurov}, Vasily and {Kravtsov}, Andrey},
        title = "{In-situ versus accreted Milky Way globular clusters: a new classification method and implications for cluster formation}",
      journal = {\mnras},
         year = 2024,
        month = feb,
       volume = {528},
       number = {2},
        pages = {3198-3216},
          doi = {10.1093/mnras/stad3920},
archivePrefix = {arXiv},
       eprint = {2309.15902},
 primaryClass = {astro-ph.GA},
       adsurl = {https://ui.adsabs.harvard.edu/abs/2024MNRAS.528.3198B}
}

@ARTICLE{Lim2025,
       author = {{Lim}, Dongwook and {Lee}, Young-Wook and {Yun}, Sol and {Lee}, Young Sun and {Chun}, Sang-Hyun and {Oh}, Heeyoung and {Lee}, Jae-Joon and {Park}, Chan and {Kim}, Sanghyuk and {Jeong}, Ueejeong and {Lee}, Hye-In and {Park}, Woojin and {Yu}, Youngsam and {Kim}, Yunjong and {Chun}, Moo-Young and {Sok Oh}, Jae and {Lee}, Sungho and {Jang}, Jeong-Gyun and {Jang}, Bi-Ho and {Seong}, Hyeon Cheol and {Kim}, Hyun-Jeong and {Brooks}, Cynthia B. and {Mace}, Gregory N. and {Lee}, Hanshin and {Good}, John M. and {Jaffe}, Daniel T. and {Kim}, Kang-Min and {Yuk}, In-Soo and {Hwang}, Narae and {Park}, Byeong-Gon and {Kim}, Hwihyun and {Chinn}, Brian and {Ramos}, Francisco and {Prado}, Pablo and {Diaz}, Ruben and {White}, John and {Tapia}, Eduardo and {Olivares}, Andres and {Oyarzun}, Valentina and {Kurz}, Emma and {Stecher}, Hawi and {Quiroz}, Carlos and {Arriagada}, Ignacio and {Hayward}, Thomas L. and {Suh}, Hyewon and {Miller}, Jen and {Xu}, Siyi and {Farina}, Emanuele Paolo and {Figura}, Charlie and {Mocnik}, Teo and {Hartman}, Zachary and {Rawlings}, Mark and {Stephens}, Andrew and {Miller}, Bryan and {Labrie}, Kathleen and {Hirst}, Paul},
        title = "{Near-Infrared Spectroscopy with IGRINS-2 for Studying Multiple Stellar Populations in Globular Clusters}",
      journal = {JKAS},
         year = 2025,
        month = apr,
       volume = {58},
        pages = {81-92},
          doi = {10.5303/JKAS.2025.58.1.81},
archivePrefix = {arXiv},
       eprint = {2504.03017},
 primaryClass = {astro-ph.GA},
       adsurl = {https://ui.adsabs.harvard.edu/abs/2025JKAS...58...81L}
}

@ARTICLE{1962King,
       author = {{King}, Ivan},
        title = "{The structure of star clusters. I. an empirical density law}",
      journal = {\aj},
         year = 1962,
        month = oct,
       volume = {67},
        pages = {471},
          doi = {10.1086/108756},
       adsurl = {https://ui.adsabs.harvard.edu/abs/1962AJ.....67..471K}
}

@ARTICLE{Van_der_Marel_2001,
       author = {{van der Marel}, Roeland P. and {Cioni}, Maria-Rosa L.},
        title = "{Magellanic Cloud Structure from Near-Infrared Surveys. I. The Viewing Angles of the Large Magellanic Cloud}",
      journal = {\aj},
         year = 2001,
        month = oct,
       volume = {122},
       number = {4},
        pages = {1807-1826},
          doi = {10.1086/323099},
archivePrefix = {arXiv},
       eprint = {astro-ph/0105339},
 primaryClass = {astro-ph},
       adsurl = {https://ui.adsabs.harvard.edu/abs/2001AJ....122.1807V}
}

@ARTICLE{Garro_Pat122Pat125Pat126,
       author = {{Garro}, E.~R. and {Minniti}, D. and {Alessi}, B. and {Patchick}, D. and {Kronberger}, M. and {Alonso-Garc{\'\i}a}, J. and {Fern{\'a}ndez-Trincado}, J.~G. and {G{\'o}mez}, M. and {Hempel}, M. and {Pullen}, J.~B. and {Saito}, R.~K. and {Ripepi}, V. and {Zelada Bacigalupo}, R.},
        title = "{Unveiling the nature of 12 new low-luminosity Galactic globular cluster candidates}",
      journal = {\aap},
         year = 2022,
        month = mar,
       volume = {659},
          eid = {A155},
        pages = {A155},
          doi = {10.1051/0004-6361/202142248},
archivePrefix = {arXiv},
       eprint = {2112.13591},
 primaryClass = {astro-ph.GA},
       adsurl = {https://ui.adsabs.harvard.edu/abs/2022A&A...659A.155G}
}

@ARTICLE{2023Garro_GaiaIngrins,
       author = {{Garro}, Elisa R. and {Fern{\'a}ndez-Trincado}, Jos{\'e} G. and {Minniti}, Dante and {Moya}, Wisthon H. and {Palma}, Tali and {Beers}, Timothy C. and {Placco}, Vinicius M. and {Barbuy}, Beatriz and {Sneden}, Chris and {Alves-Brito}, Alan and {Dias}, Bruno and {Af{\c{s}}ar}, Melike and {Frelijj}, Heinz and {Lane}, Richard R.},
        title = "{Gaia-IGRINS synergy: Orbits of newly identified Milky Way star clusters}",
      journal = {\aap},
         year = 2023,
        month = jan,
       volume = {669},
          eid = {A136},
        pages = {A136},
          doi = {10.1051/0004-6361/202245119},
archivePrefix = {arXiv},
       eprint = {2212.02337},
 primaryClass = {astro-ph.GA},
       adsurl = {https://ui.adsabs.harvard.edu/abs/2023A&A...669A.136G}
}

@ARTICLE{2024_Garro,
       author = {{Garro}, E.~R. and {Minniti}, D. and {Fern{\'a}ndez-Trincado}, J.~G.},
        title = "{Over 200 globular clusters in the Milky Way and still none with super-Solar metallicity}",
      journal = {\aap},
         year = 2024,
        month = jul,
       volume = {687},
          eid = {A214},
        pages = {A214},
          doi = {10.1051/0004-6361/202347389},
archivePrefix = {arXiv},
       eprint = {2405.05055},
 primaryClass = {astro-ph.GA},
       adsurl = {https://ui.adsabs.harvard.edu/abs/2024A&A...687A.214G}
}

@ARTICLE{2017_Alonso-Garcia,
       author = {{Alonso-Garc{\'\i}a}, Javier and {Minniti}, Dante and {Catelan}, M{\'a}rcio and {Contreras Ramos}, Rodrigo and {Gonzalez}, Oscar A. and {Hempel}, Maren and {Lucas}, Philip W. and {Saito}, Roberto K. and {Valenti}, Elena and {Zoccali}, Manuela},
        title = "{Extinction Ratios in the Inner Galaxy as Revealed by the VVV Survey}",
      journal = {\apjl},
         year = 2017,
        month = nov,
       volume = {849},
       number = {1},
          eid = {L13},
        pages = {L13},
          doi = {10.3847/2041-8213/aa92c3},
archivePrefix = {arXiv},
       eprint = {1710.04854},
 primaryClass = {astro-ph.GA},
       adsurl = {https://ui.adsabs.harvard.edu/abs/2017ApJ...849L..13A}
}

@ARTICLE{2024_Petralia,
       author = {{Petralia}, Ilaria and {Minniti}, Dante and {Fern{\'a}ndez-Trincado}, Jos{\'e} G. and {Lane}, Richard R. and {Schiavon}, Ricardo P.},
        title = "{Signature of systemic rotation in 21 galactic globular clusters from APOGEE-2}",
      journal = {\aap},
         year = 2024,
        month = aug,
       volume = {688},
          eid = {A92},
        pages = {A92},
          doi = {10.1051/0004-6361/202347550},
archivePrefix = {arXiv},
       eprint = {2404.10902},
 primaryClass = {astro-ph.GA},
       adsurl = {https://ui.adsabs.harvard.edu/abs/2024A&A...688A..92P}
}

@ARTICLE{2004_Pichardo,
       author = {{Pichardo}, B{\'a}rbara and {Martos}, Marco and {Moreno}, Edmundo},
        title = "{Models for the Gravitational Field of the Galactic Bar: An Application to Stellar Orbits in the Galactic Plane and Orbits of Some Globular Clusters}",
      journal = {\apj},
         year = 2004,
        month = jul,
       volume = {609},
       number = {1},
        pages = {144-165},
          doi = {10.1086/421008},
archivePrefix = {arXiv},
       eprint = {astro-ph/0402340},
 primaryClass = {astro-ph},
       adsurl = {https://ui.adsabs.harvard.edu/abs/2004ApJ...609..144P}
}

@ARTICLE{2022_Moreno,
       author = {{Moreno}, Edmundo and {Fern{\'a}ndez-Trincado}, Jos{\'e} G. and {P{\'e}rez-Villegas}, Angeles and {Chaves-Velasquez}, Leonardo and {Schuster}, William J.},
        title = "{Orbits of globular clusters computed with dynamical friction in the Galactic anisotropic velocity dispersion field}",
      journal = {\mnras},
         year = 2022,
        month = mar,
       volume = {510},
       number = {4},
        pages = {5945-5962},
          doi = {10.1093/mnras/stab3724},
archivePrefix = {arXiv},
       eprint = {2112.11589},
 primaryClass = {astro-ph.GA},
       adsurl = {https://ui.adsabs.harvard.edu/abs/2022MNRAS.510.5945M}
}

\appendix
\section{Data and radial velocities of the analysed stars}\label{App_table_1}
In this appendix, Table \ref{Table2} presents the data and the {RVs}
derived for the 33 stars analysed in this study.

\renewcommand{\arraystretch}{1.5} 
\begin{sidewaystable*}[h!]
\begin{scriptsize}
    \centering
    \captionsetup{font=footnotesize}
    \caption{\scriptsize Summary of the 33 stars analysed in this work.}
    \resizebox{\textheight}{!}{%
    \begin{tabularx}{\textwidth}{ccccccccccccc}
        \hline
        ID Gaia DR3 & Cluster& RA & Dec & J band & Date of Observation & Time of observation& Heliocentric Correction & \multicolumn{2}{c}{S/N per pixel} & Orders&WINERED RVs & Gaia RVs \\
         &&  [deg] & [deg] & [mag] & [UT]& [UT]& [km/s]& @Y & @J &&  [km/s] & [km/s] \\
        \hline
        5534906319798217984 & \multirow{7}*{CWNU~4193}&
        121.111750 &$-$38.887639& 13.507 &2023-11-02&09:12:06.883 to 09:21:05.889&16.14&$\sim3$&$\sim5$ &47-46&{137.19}
        $\pm$ 0.82 & - \\
        5540910031242716928 &&  121.172621 & $-$38.916016 & 12.423 &2024-04-19&00:57:19.539 to 01:33:36.565&$-$15.27&$\sim20$ &$\sim28$&55-54-53& {136.36}
        $\pm$ 0.65 & - \\
        5534905976200827776 &&  121.150405 & $-$38.918991 & 11.933 &2024-04-19&01:43:15.571 to 02:12:18.591&$-$14.97& $\sim80$&$\sim98$&55-54-53&{138.09}
        $\pm$ 0.66 & 140.15 $\pm$ 3.77\\
        5540909893796469632&  &121.210278&$-$38.930618&12.995&2025-02-09&00:37:25.525 to 01:06:43.546&$-$0.36&$\sim20$&$\sim25$&55-54-53& {71.56}
        $\pm$ 1.05 & - \\
        5534905078543822208&  &121.137692&$-$38.962906&13.806&2025-02-09&01:25:21.559 to 04:45:30.698&$-$0.40&$\sim13$&$\sim13$&55-54-53& {136.58}
        $\pm$ 0.71 & - \\
        5534905082848371712& &121.131068&$-$38.955254&14.392&2025-02-09&02:24:39.600 to 03:26:17.643&$-$0.52&$\sim12$&$\sim17$&55-54-53&{134.91}
        $\pm$ 0.84 & - \\
        5540910164378603776& &121.190785&$-$38.900639&14.439&2025-02-09&03:42:37.654 to 04:16:54.678&$-$0.60&$\sim12$&$\sim16$&55-54-53& {136.53}
        $\pm$ 0.76 & - \\
        \hline        
        5881863772347874816 & \multirow{7}*{FSR~1700}&  234.582914 & $-$59.233246 & 11.196&2024-04-19&05:30:55.729 to 05:44:56.739&14.51& $\sim130$ &$\sim162$&55-54-53& {5.59}
        $\pm$ 0.62 & 7.87 $\pm$ 2.82 \\
        5881868582711187072 & & 234.686805 & $-$59.257389 & 10.938&2024-04-19&05:51:41.744 to 06:01:15.750&14.62& $\sim96$&$\sim136$&55-54-53&{3.34}
        $\pm$ 0.68 & 6.71 $\pm$ 1.86 \\
        5881868587017262848 && 234.681311&$-$59.25301 &13.487&2025-02-09&09:07:36.880 to 09:43:58.905&22.61&$\sim5$&$\sim8$&47-46&{7.09}
        $\pm$ 0.68 & - \\
        5881868621376996480 &&  234.712388& $-$59.246178&13.669&2025-02-09&08:50:29.868 to 09:00:59.875&-&$\sim4$&$\sim5$&-&- & - \\
        5881868449635849984 & &234.758337&$-$59.252148&13.765&2025-02-09&08:32:51.856 to 08:43:19.863&22.64&$\sim8$&$\sim10$&47-46&{6.11}
        $\pm$ 0.53 & - \\
        5881862471040696320 & &234.664319&$-$59.305744&13.807&2025-02-09&08:15:24.844 to 08:26:11.851&22.64&$\sim7$&$\sim11$&55-54-53&{-25.89}
        $\pm$ 0.64 & - \\
        5881868445272228352 & &234.770252&$-$59.256763&13.85&2025-02-09&07:54:48.829 to 08:06:03.837&22.66&$\sim6$&$\sim9$&47-46&{6.24}
        $\pm$ 0.55 & - \\
        \hline         
        4143904951111772928  & \multirow{3}*{Garro~02}&271.48972&$-$17.70828&10.429&2024-09-10&00:55:14.538 to 01:04:41.544&$-$28.89&$\sim113$&$\sim197$&55-54-53&{167.17}
        $\pm$ 0.89 & 163.27 $\pm$ 3.38 \\
        4143907768610411904& &271.4411&$-$17.698725&10.529&2024-09-10&01:09:23.548 to 01:20:26.555&$-$28.92&$\sim12$&$\sim7$&55-54-53&{169.02}
        $\pm$ 0.78 & 166.76 $\pm$ 4.93\\
        4143904710593625088& &271.45694&$-$17.727825&10.784&2024-09-10&01:26:38.560 to 01:39:08.568&$-$28.95&$\sim110$&$\sim190$&55-54-53&{170.19}
        $\pm$ 0.63 & 168.19 $\pm$ 3.50 \\
        \hline
        4153507187997149312& \multirow{4}*{Patchick~98}& 274.60782&$-$12.558493&9.876&2024-09-09&00:24:02.516 to 00:35:12.524&$-$27.89&$\sim160$&$\sim290$&55-54-53&{31.38}
        $\pm$ 0.45 & - \\
        4153509803645216896& & 274.64096&$-$12.48168&10.404&2024-09-12&00:01:08.500 to 00:18:15.512&$-$28.28&$\sim165$&$\sim280$&55-54-53&{14.47}
        $\pm$ 0.66 & - \\
        4153508394895997312&& 274.56564&$-$12.516385&10.464&2024-09-08&23:42:20.487 to 23:56:30.497&$-$27.82&$\sim145$&$\sim230$&55-54-53&{-32.50}
        $\pm$ 0.69 & - \\
        4153507329743961600&& 274.6337&$-$12.539847&10.505&2024-09-09&00:04:13.502 to 00:15:16.510&$-$27.86&$\sim105$&$\sim180$&55-54-53&{76.35}
        $\pm$ 0.60 & 85.65 $\pm$ 7.21\\
        \hline
        5962720315656305792& \multirow{3}*{FSR~1767}&263.9093&$-$36.355007&10.155&2024-09-09&03:23:23.641 to 03:34:27.648&$-$28.89&$\sim208$&$\sim280$&55-54-53&{-64.70}
        $\pm$ 0.63 & - \\
        5962723408031884672& &263.9571&$-$36.3286&10.202&2024-09-10&00:32:56.522 to 00:43:59.530&$-$28.78&$\sim148$&$\sim230$&55-54-53&{59.08}
        $\pm$ 0.58 & - \\
        5962717257598211072&& 263.91458&$-$36.389305&10.615&2024-09-09&03:47:53.658 to 04:06:59.671&$-$28.90&$\sim74$&$\sim103$&55-54-53&{16.80}
        $\pm$ 0.38 & - \\
        \hline   
        4154940199552504960 &\multirow{2}*{Mercer~08}& 277.207835&$-$10.935265&11.512&2024-09-10&02:01:28.584 to 02:06:35.587&$-$27.58&$\sim12$&$\sim35$&55-54-53&{93.89}
        $\pm$ 0.35 & - \\
        4154940165180403712 & &277.192694&$-$10.941928&11.83&2024-09-10&02:17:17.595 to 02:22:30.598&$-$27.60&$\sim9$&$\sim29$&55-54-53&{9.81}
        $\pm$ 0.42 & - \\
        \hline
        5858163455614441728 &\multirow{7}*{BH~140 }& 193.31467&$-$67.104256&8.908&2025-02-09&07:27:37.810 to 07:31:31.813&17.75&$\sim153$&$\sim175$&55-54-53&{92.47}
        $\pm$ 0.49 & 89.16 $\pm$ 0.34 \\
        5858115321965911808 &&193.28607& $-$67.17836&9.278&2025-02-09&07:18:03.804 to 07:22:23.807&17.72&$\sim89$&$\sim133$&55-54-53&{90.89}
        $\pm$ 0.95 & 90.50 $\pm$ 0.29\\
        5858112023430585216 &&  193.31717&$-$67.230034&9.384&2025-02-09&07:03:41.794 to 07:12:54.800&17.72&$\sim107$&$\sim148$&55-54-53&{102.81}
        $\pm$ 1.34 & 90.71 $\pm$ 2.48\\
        5858113603978703872 && 193.49342& $-$67.18431&10.527&2025-02-09&06:50:39.785 to 06:58:51.790&17.76&$\sim57$&$\sim84$&55-54-53&{85.61}
        $\pm$ 1.07 & 100.47 $\pm$ 5.17\\
        5858110992638378240 & &193.43056&$-$67.26718&11.121&2025-02-09&06:36:20.775 to 06:44:42.781&17.74&$\sim45$&$\sim60$&55-54-53&{91.64}
        $\pm$ 1.06 & 88.81 $\pm$ 1.83\\
        5858116936873419136 && 193.418508&$-$67.148735&13.065&2025-02-09&06:02:18.751 to 06:28:49.770&17.79&$\sim19$&$\sim14$&54-53&{88.93}
        $\pm$ 1.00 & - \\
        5858116657690170624 && 193.427354&$-$67.165909&13.148&2025-02-09&05:16:10.719 to 05:48:53.742&17.81&$\sim10$&$\sim9$&54-53-47&{89.71}
        $\pm$ 1.13 & - \\
        \hline
    \end{tabularx}
    }
   \tablefoot{\scriptsize The target names, the membership cluster and their coordinates are listed in columns 1 - 4. The J-band magnitude for each star is listed in the fifth column.
   {The date of observation of the stars is shown in the sixth column, while the time of observation in UT is in the seventh column. The eighth column shows the heliocentric correction for all stars. All 33 were observed using the WINERED spectrograph attached to the Magellan Clay 6.5 m telescope at Las Campanas Observatory (LCO), Chile. Hereafter, the observatory parameters for LCO from rvcorrect task within IRAF: latitude = -29:0.2, longitude = 70:42.1 and altitude = 2282.}
   {Signal-to-noise ratio per pixel of the observed WINERED spectra are shown in the {ninth} column. S/N @Y and S/N @J {are shown in the {ninth} column and} were estimated using the echelle order 54 and 46, respectively. The S/N value is computed from the S/N values within the spectral order and represents an approximate average. The tenth} column shows the echelle orders used to derive the {RV.}
   The {eleventh} column shows the {RV}
   values after the heliocentric correction for each star. {Finally, the last column shows the {RV values} from Gaia’s {RV}
   spectrometer.} }
    \label{Table2}
 \end{scriptsize}   
\end{sidewaystable*}

\section{Estimation of intrinsic velocity dispersion}\label{App_intrisic_dispersion}In this appendix, we report the steps followed to derive the intrinsic velocity dispersion {($\sigma_{int}$)} of each star cluster.\\
First, for each cluster, we calculated the weighted mean of the {RVs}:\\
\begin{equation}
    \mathrm{\overline{\mathrm{RV}}_{w} = \frac{\sum\limits_{n=1}^N\frac{\mathrm{RV}_{n}}{\delta_{\mathrm{RV}_{n}}^2}}{\sum\limits_{n=1}^N\frac{1}{\delta_{RV_{n}}^2}}}.
\end{equation}
Then, we computed the observed variance,\\
\begin{equation}
    \sigma_{obs}^2 = \mathrm{\frac{\sum\limits_{n=1}^N\left(\frac{(\mathrm{RV}_{n}-\overline{\mathrm{RV}}_{w})^{2}}{\delta^2_{\mathrm{RV}_{n}}}\right)}{\sum\limits_{n=1}^N\left(\frac{1}{\delta^2_{\mathrm{RV}_{n}}}\right)}},
\end{equation}

\noindent and the mean variance of the measurement errors:\\
\begin{equation}
    \sigma_{err}^2 = \mathrm{\mathrm{\frac{1}{N}}\sum\limits_{n=1}^N(\delta_{\mathrm{RV}_{n}})^2}.
\end{equation}

\noindent Finally, we derived the {$\sigma_{int}$}
:\\
\begin{equation}
    \sigma_{int} = \sqrt{\sigma_{obs}^2 - \sigma_{err}^2}.
\end{equation}

\noindent For each of the clusters analysed in this study, we derived the {$\sigma_{int}$}
using the
N available {RV}
measurements for each stellar system.\\
The uncertainty on the intrinsic velocity dispersion {($\delta_{\sigma_{int}}$)} were derived through the standard error propagation\\
\begin{equation}
    \delta_{\sigma_{int}} = \frac{1}{\sigma_{int}} \sqrt{(\sigma_{obs}\cdot\delta_{\sigma_{obs}})^2+(\sigma_{err}\cdot\delta_{\sigma_{err}})^2},
\end{equation}\\
\noindent where
\begin{equation}
    \delta_{\sigma_{obs}} = \frac{\sigma_{obs}}{\sqrt{\mathrm{2(N-1)}}}
 \quad \text{and} \quad
    \delta_{\sigma_{err}} = \frac{std(\delta_{\mathrm{RV}_{n}})}{\sqrt{\mathrm{N}}}.
\end{equation}
The results derived from this appendix are presented in Table \ref{table4}.\\

\section{ Qualitative analysis for FSR~1767, Patchick~98, and Mercer~08}\label{speculative_analisi}

{ In this appendix, we present a qualitative analysis for the clusters FSR~1767, Patchick~98, and Mercer~08. The results emphasise the necessity of follow-up spectroscopy for more candidates.
Even so, we provides some general considerations and speculative predictions of the orbital paths for these clusters in basis of the measured RVs of their member star candidates.}\\

{Our first considerations concern {RVs}
of the candidate member stars.
In particular, for the clusters FSR 1767 and Patchick 98, the velocities differ by several tens of km/s, which, according to the 3$\sigma$ rejection criterion, suggest that not all candidate stars may belong to the cluster. Nevertheless, it should be emphasised that in star clusters, as in the case of GCs, the {RVs} 
of member stars often spans a broader range.
Hence, the candidate stars analysed in this work may be members of the clusters.
Since these clusters have never been studied from a kinematic perspective, the expected ranges of {RVs}
of their member stars are unknown. For this reason, we adopted the 3$\sigma$ rejection criterion to perform a more reliable analysis.
Further observations of candidate members will be crucial to better constrain the {RV}
ranges of stars belonging to these clusters and to assess more reliably the membership of individual stars.\\
Another relevant point is that several candidate stars of a given cluster exhibit similar absolute {radial velocities}, differing by only a few kilometres per second, but with opposite signs. This does not necessarily exclude them from membership.
Indeed, opposite velocity signs may simply indicate that some stars are blueshifted while others are redshifted, which could be a signature of systemic rotation in the clusters.
To test this hypothesis, it is necessary to obtain {RVs}
for a sufficiently large sample of members (at least $\sim$ 60 stars, spatially uniformly distributed). 
Such data would allow us to evaluate the presence of rotation and to determine whether stars with opposite signs indeed belong to the same star system. For further details on these first considerations, see \cite{2024_Petralia}.}\\

{In the following, on the other hand, we present the possible orbital paths for the three clusters.}\\
{ 
As shown in Table \ref{tab:Orbital_results}, FSR~1767 is likely confined in a bulge-like orbital configuration, which lies in an in-plane orbit with high eccentricity $\gtrsim$ {0.89} and low vertical ($Z_{\rm max} \lesssim $ 1.2 kpc) excursions from the Galactic plane, and {$r_{apo}$}
below 3.1 kpc, a {$r_{peri}$ between 0.04 and 0.21 kpc.}
This cluster is going inside and outside of the bar in the Galactic plane, but with not bar-shape orbit, which means that this object is not trapped by the bar. The effect produced by different {$\Omega_{\rm bar}$}
on this object reveal a prograde orbit for negative values of RV, while positive values of RV might characterise a chaotic behaviour with orbits that change their sense of motion from prograde to retrograde {{(see e.g. \citealt{2004_Pichardo}, \citealt{Perez-Villegas2018}, \citealt{2022_Moreno})}}. If some of the measured RVs are confirmed, it would be feasible to describes to FSR~1767 as a cluster that lives in the inner bulge region, likely confined to a chaotic orbit affected by the bar motion. 
}\\
{ 
Patchick~98 is also likely confined in a prograde bulge-like orbital configuration, which lies in an in-plane orbit with relatively high eccentricity between 0.49 {and} 0.81 and very low vertical ($Z_{\rm max} \lesssim $ 0.3 kpc) excursions from the Galactic plane, and {$r_{apo}$}
below $\sim$ 4.1 kpc, a {$r_{peri}$}
between {0.38 and 1.38 kpc}. This cluster is also going inside and outside of the bar in the Galactic plane, but with not bar-shape orbit, which means that this object is likely not trapped by the bar. If the measured RVs are confirmed, it would be feasible to describes to Patchick~98 as a cluster that also lives in the inner bulge region. 
}\\
{ 
Finally, Mercer~08 is confined in a prograde orbital configuration, which lies in a low eccentricity between 0.18 to 0.44 with very low vertical ($Z_{\rm max}$ {$\sim $} 0.04 kpc) excursions from the Galactic plane, with a {$r_{peri}$}
between {2 and 4} kpc, and {$r_{apo}$}
between 5.0 and 6.3 kpc, which cross multiple times the {CR}.
If the measured RVs are confirmed, it would be feasible to describes to Mercer~08 as a cluster assembling an orbit more consistent with the Galactic disc.
}

\renewcommand{\arraystretch}{1.5} 
\begin{table*}
\begin{small}
 \centering
 \captionsetup{font=footnotesize}
 \caption{{ Orbital elements.}}
 \begin{tabular}{cccccccc} 
 \hline
 \noalign{\smallskip}
 IDs    & $\Omega_{bar}$&  $r_{peri}$  & $r_{apo}$ & Z$_{\rm max}$ & e & Orbital sense \\ 
  &  & [kpc] & [kpc] & [kpc] & & \\

 \noalign{\smallskip}
 \hline
 \noalign{\smallskip}

 CWNU~4193 &   \multirow{13}*{31 km s$^{-1}$ kpc$^{-1}$ }&
 11.90$\pm$2.41 & 17.13$\pm$0.92 & 1.29$\pm$0.66 & 0.18$\pm$0.09 & Prograde \\
 FSR~1700 && 3.50$\pm$0.87 & 6.43$\pm$0.56 & 0.63$\pm$0.09  & 0.29$\pm$0.07 & Prograde \\
 Garro~02  && 0.97$\pm$0.28 & 4.15$\pm$1.17 & 0.81$\pm$0.59 & 0.62$\pm$0.05 & Prograde \\
 BH~140  && 1.99$\pm$0.50 & 10.02$\pm$2.61 & 1.52$\pm$0.58 & 0.67$\pm$0.02 & Prograde \\
 5962720315656305792 (FSR~1767) && 0.17$\pm$0.23 & 2.97$\pm$0.89  & 0.56$\pm$0.27 & 0.89$\pm$0.11 & Prograde \\
 5962723408031884672 (FSR~1767) && 0.06$\pm$0.16 & 2.98$\pm$1.00  & 0.78$\pm$0.52 & 0.96$\pm$0.08 & Prograde-Retrograde \\
 5962717257598211072 (FSR~1767) && 0.08$\pm$0.18 & 2.90$\pm$0.95 & 0.60$\pm$0.40 & 0.95$\pm$0.09 & Prograde-Retrograde\\
 4153507187997149312 (Patchick~98) && 1.00$\pm$0.85 & 3.68$\pm$0.69 & 0.27$\pm$0.11 & 0.56$\pm$0.23 & Prograde \\
 4153509803645216896 (Patchick~98) && 0.79$\pm$0.75 & 3.68$\pm$0.70 & 0.23$\pm$0.10 & 0.65$\pm$0.21 & Prograde \\
 4153508394895997312 (Patchick~98) && 0.64$\pm$0.64 & 3.62$\pm$0.70 & 0.19$\pm$0.08 & 0.70$\pm$0.20 & Prograde \\
 4153507329743961600 (Patchick~98) && 1.38$\pm$0.78 & 4.06$\pm$0.84 & 0.25$\pm$0.10 & 0.49$\pm$0.15 & Prograde \\
 4154940199552504960 (Mercer~08) && 3.99$\pm$0.59 & 5.55$\pm$1.20 & 0.04$\pm$0.03 & 0.18$\pm$0.05 & Prograde \\
 4154940165180403712 (Mercer~08) && 2.26$\pm$1.23 & 5.07$\pm$0.82  & 0.04$\pm$0.04 & 0.38$\pm$0.16 & Prograde \\
 \noalign{\smallskip}
 \hline
 CWNU~4193 &   \multirow{13}*{41 km s$^{-1}$ kpc$^{-1}$ }&
 11.90$\pm$2.40 & 17.08$\pm$1.02 & 1.29$\pm$0.66 &  0.18$\pm$0.08 & Prograde \\
 FSR~1700 && 3.49$\pm$0.67 & 7.13$\pm$0.83 & 0.66$\pm$0.11 & 0.35$\pm$0.06 & Prograde \\
 Garro~02  && 0.98$\pm$0.37 & 4.33$\pm$1.20 & 0.87$\pm$0.29 & 0.61$\pm$0.08 & Prograde \\
 BH~140  && 1.80$\pm$0.59 & 9.70$\pm$2.56 & 1.05$\pm$0.72 & 0.68$\pm$0.04 & Prograde \\
 5962720315656305792 (FSR~1767) && 0.17$\pm$0.22  & 2.96$\pm$0.88 & 0.69$\pm$0.54 & 0.89$\pm$0.10 & Prograde \\
 5962723408031884672 (FSR~1767) && 0.04$\pm$0.13 & 3.03$\pm$1.01 & 1.13$\pm$0.61 & 0.97$\pm$0.06 & Prograde-Retrograde \\
 5962717257598211072 (FSR~1767) && 0.05$\pm$0.15 & 2.89$\pm$0.96  & 0.86$\pm$0.60 & 0.96$\pm$0.07 & Prograde-Retrograde \\
 4153507187997149312 (Patchick~98) && 0.91$\pm$0.44 & 3.70$\pm$0.69 & 0.26$\pm$0.11 & 0.60$\pm$0.12 & Prograde \\
 4153509803645216896 (Patchick~98) && 0.80$\pm$0.43 & 3.69$\pm$0.69 & 0.23$\pm$0.10 & 0.64$\pm$0.13 & Prograde \\
 4153508394895997312 (Patchick~98) && 0.64$\pm$0.40 & 3.65$\pm$0.68  & 0.19$\pm$0.09 & 0.70$\pm$0.12 & Prograde \\
 4153507329743961600 (Patchick~98) && 1.17$\pm$0.61 & 4.08$\pm$0.87 & 0.23$\pm$0.08 & 0.52$\pm$0.10 & Prograde \\
 4154940199552504960 (Mercer~08) && 3.75$\pm$0.74 & 5.51$\pm$1.64 & 0.04$\pm$0.03 & 0.21$\pm$0.05 & Prograde \\
 4154940165180403712 (Mercer~08) && 2.27$\pm$1.24 & 5.09$\pm$0.81 & 0.04$\pm$0.03 & 0.38$\pm$0.19 & Prograde \\   
 \noalign{\smallskip}
 \hline
 CWNU~4193 &\multirow{13}*{51 km s$^{-1}$ kpc$^{-1}$ }&
 11.90$\pm$2.42 & 17.10$\pm$0.99 & 1.29$\pm$0.65 & 0.18$\pm$0.09 & Prograde \\
 FSR~1700 && 2.65$\pm$0.89 & 6.39$\pm$0.42 & 0.60$\pm$0.09 & 0.40$\pm$0.10 & Prograde \\
 Garro~02  && 1.03$\pm$0.30 & 4.34$\pm$1.12 & 1.05$\pm$0.47 & 0.63$\pm$0.08 & Prograde\\
 BH~140  && 1.95$\pm$0.58 & 9.88$\pm$2.71 & 1.09$\pm$0.79 & 0.69$\pm$0.04 & Prograde \\
 5962720315656305792 (FSR~1767) && 0.21$\pm$0.27 & 3.02$\pm$0.92 & 0.84$\pm$0.46 & 0.87$\pm$0.11 &  Prograde\\
 5962723408031884672 (FSR~1767) && 0.04$\pm$0.15 & 3.07$\pm$1.05 & 1.16$\pm$0.52 & 0.97$\pm$0.06 & Prograde-Retrograde\\
 5962717257598211072 (FSR~1767) && 0.06$\pm$0.23 & 2.94$\pm$1.04 & 0.97$\pm$0.49 & 0.96$\pm$0.10 & Prograde-Retrograde \\
 4153507187997149312 (Patchick~98) && 0.62$\pm$0.71 & 3.81$\pm$0.70 & 0.26$\pm$0.08 & 0.70$\pm$0.19 & Prograde\\
 4153509803645216896 (Patchick~98) && 0.56$\pm$0.68 & 3.85$\pm$0.70 & 0.22$\pm$0.07  & 0.73$\pm$0.19 & Prograde \\
 4153508394895997312 (Patchick~98) && 0.38$\pm$0.67 & 3.74$\pm$0.69 & 0.19$\pm$0.07 & 0.81$\pm$0.20 & Prograde \\
 4153507329743961600 (Patchick~98) && 1.28$\pm$0.66 & 4.09$\pm$0.92 & 0.22$\pm$0.08 & 0.51$\pm$0.14 & Prograde \\
 4154940199552504960 (Mercer~08) && 3.05$\pm$0.99 & 6.28$\pm$1.36 & 0.04$\pm$0.04 & 0.24$\pm$0.07 & Prograde\\
 4154940165180403712 (Mercer~08) && 2.00$\pm$1.20 & 5.13$\pm$1.19 & 0.04$\pm$0.03 & 0.44$\pm$0.12 & Prograde \\
 \noalign{\smallskip}
 \hline
 \end{tabular}
 
 \tablefoot{{\scriptsize The cluster and star names are listed in the first column{, while the $\Omega_{\rm bar}$ is presented in the second column.} The {$r_{peri}$ and the $r_{apo}$} are shown in the {third and fourth} columns, respectively. In the {fifth column there is the Z$_{\rm max}$}, while in the sixth column there is the eccentricity. Finally, in the last columns, the orbital sense is reported. The average value of the orbital elements was found for one million realisations, with uncertainty ranges given by the 16$^{\rm th}$ and 84$^{\rm th}$ percentile values. The errors provided in each column are computed as $\Delta$ $=$ 0.5 $\times$ (84$^{\rm th}$ percentile $-$ 16$^{\rm th}$ percentile). }}\label{tab:Orbital_results}

\end{small}
\end{table*}    

\section{Calculation of the tidal radius of the BH 140 cluster}\label{app:tidal_radius_bh140}
\begin{figure}[H]
   
    \centering
    \includegraphics[width=0.9\linewidth]{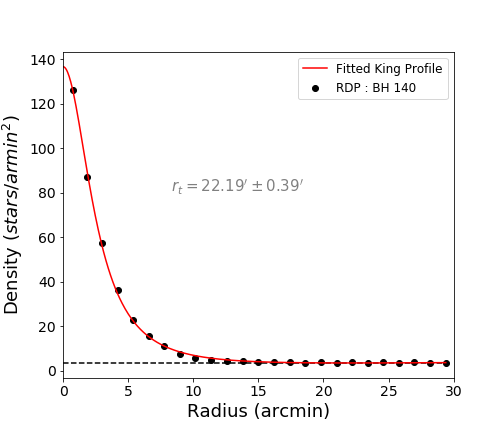}
    \caption{Radial density profile of the BH 140 cluster. The black points correspond to the cluster density profile, which corresponds to the number of stars per unit area within an annular region. The red line shows the best-fit King model profile (\citealt{1962King}). The constant background density is marked with the dashed line.}
    \label{fig:king_profile}
\end{figure}

In this appendix, we describe the method to derive the tidal radius ($r_{t}$) of the BH 140 cluster, starting from its photometric Gaia DR3 data. To determine this parameter, we computed the radial density profile (RDP)
and employed the King model (\citealt{1962King}).
For this analysis, we selected all stars within a 30-arcminute radius from the cluster centre and those with proper motions in right ascension ($\mu_\alpha^{*}$)\footnote{{By convention, the mean proper motion in right ascension is given as $\mu_{\alpha}^{*}= \mu_{\alpha} \cdot cos\delta$}} and declination ($\mu_\delta$) within the following ranges: $ \mu_\alpha^{*} \in [\overline{\mu_{\alpha^{*}}}-5\cdot\delta_{\overline{\mu_{\alpha^{*}}}}, \overline{\mu_{\alpha^{*}}}+5\cdot\delta_{\overline{\mu_{\alpha^{*}}}}] $ and  
$\mu_\delta \in [\overline{\mu_{\delta}}-5\cdot\delta_{\overline{\mu_{\delta}}}, \overline{\mu_{\delta}}+5\cdot\delta_{\overline{\mu_{\delta}}}]$;
where $\overline{\mu_\alpha^{*}}$ and $\delta_{\overline{\mu_\alpha^{*}}}$ are the mean proper motion in right ascension of the cluster and its uncertainty, while $\overline{\mu_\delta}$ and $\delta_{\overline{\mu_\delta}}$ are the mean proper motion in declination
of the cluster and its uncertainty. The values of the proper motions and their uncertainties\footnote{The value of the uncertainties reported was multiplied by the square root of the number of stars analysed in the reference literature.} along with the cluster’s centre coordinates were adopted from \citet{Vasiliev-Baumgardt+21}. With this selection, we were able to consider a sufficiently large region (larger than the {$r_{t}$})
and to select an adequate number of stars within it, including both member and field stars, in order to fit the density profile of this cluster. The following strategy was employed to construct the RDP. First, the Galactic coordinates of the stars were converted into Cartesian coordinates using the transformation equations from \citet{Van_der_Marel_2001}. Then, we divided the sample into 25 circular different annuli with increasing radii, each with a width of 0.02 deg (1.2 arcmin) and all centred on the cluster centre. The number density per bin was calculated as the number of stars in the bin divided by the respective area. The RDP of the cluster is plotted as a function of the mean distance of the circular annulus from the cluster centre to the number density of stars in the corresponding annuli. Finally, we adopted the King model to fit the cluster density profile. We used a chi-square method to derive the best-fitting model parameters and the standard errors. The best-fit King model gives the $r_{t}$ of the cluster (see, Fig. \ref{fig:king_profile}). As a result, we found that the {$r_{t}$} 
of the BH 140 cluster, derived from the King profile, is 22.19$'$ $\pm$ 0.39$'$, which corresponds to 31.0 $\pm$ 1.7 pc.

\end{document}